\documentclass[12pt,preprint]{aastex}
\shorttitle{The Star Formation History of M31}

\received{2006 February 2}
\accepted{2006 March 27}
\begin{document}

\title{The Star Formation Histories of the Bulge and Disk of M31 from Gemini North+NIRI/Altair and HST/NICMOS\footnote{Based on observations obtained at the
    Gemini Observatory, which is operated by the Association of
    Universities for Research in Astronomy, Inc., under a cooperative
    agreement with the NSF on behalf of the Gemini partnership: the
    National Science Foundation (United States), the Particle Physics
    and Astronomy Research Council (United Kingdom), the National
    Research Council (Canada), CONICYT (Chile), the Australian
    Research Council (Australia), CNPq (Brazil) and CONICET
    (Argentina).  Based on observations made with the NASA/ESA Hubble Space Telescope, obtained at the Space Telescope Science Institute, which is operated by the Association of Universities for Research in Astronomy, Inc., under NASA contract NAS 5-26555.}
}

\author{Knut A.G. Olsen, Robert D. Blum}\affil{National Optical Astronomy Observatory,
CTIO, Casilla 603, La Serena, Chile}\email{kolsen@ctio.noao.edu, rblum@ctio.noao.edu}

\author{Andrew W. Stephens}\affil{Gemini Observatory,
 670 N. A'ohoku Place, Hilo, Hawaii 96720}\email{stephens@gemini.edu}

\author{Tim J. Davidge}\affil{Herzberg Institute of Astrophysics, National Research Council of Canada, 5071 West Saanich Road, Victoria, BC V9E 2E7, Canada}\email{Tim.Davidge@hia-iha.nrc-cnrc.gc.ca}

\author{Philip Massey}\affil{Lowell Observatory, 1400 West Mars Hill Road, Flagstaff, AZ 86001; Phil.Massey@lowell.edu}

\author{Stephen E. Strom}\affil{National Optical Astronomy Observatory, 950 N. Cherry Ave., Tucson, AZ 85719}\email{sstrom@noao.edu}

\author{Fran\c{c}ois Rigaut}\affil{Gemini Observatory, 670 N. A'ohoku Place,
Hilo, Hawaii, 96720}\email{frigaut@gemini.edu}

\begin{abstract}

We discuss $H$ and $K$ observations of three fields in the bulge
and disk of M31 obtained with the Altair adaptive optics system and
NIRI instrument on Gemini North. These are the highest resolution and
deepest near-infrared observations obtained to date of the inner
regions of M31, and demonstrate the promise of ground-based adaptive
optics for studying the crowded regions of nearby galaxies.  We have
combined our observations with previously published {\it HST/NICMOS}
observations of nine M31 fields and have derived the coarse star
formation histories of M31's bulge and inner disk.  From fits to the $M_K$
luminosity functions, we find the stellar population mix to be
dominated by old, nearly solar-metallicity stars.  The old
populations, which we define as having age $\gtrsim$6 Gyr, indeed dominate
the star formation histories at all radii independent of the relative
contributions of bulge and disk stars.  Although all of our fields contain some
bulge contribution, our results suggest that there is no age difference between the bulge and disk to the limit of our precision.

\end{abstract}

\section{INTRODUCTION}
The problem of galaxy formation occupies a central role in
astrophysics, as it connects the study of cosmology to that of star
formation and the dispersal of chemical elements.  Recent observations
of young galaxies at high redshift show that many massive galaxies
were already in place by $z\sim$2, challenging some models of galaxy
formation (e.g. Glazebrook et al.\ 2004).  However, the observations
contain insufficient information to determine whether these galaxies
are in the process of assembling disks or spheroids.
Indeed, a more sensitive test of galaxy
formation physics appears to be to examine the star formation
histories of galaxy disks compared to those of bulges.  Two recent
simulations of forming disk galaxies, differing mainly in their
treatment of stellar feedback, produce wildly different results (Abadi
et al. 2003, Robertson et al. 2004).  On one hand, the disk galaxy
found in the Abadi et al. simulation has a massive old spheroid and a
much younger, $\sim$4 Gyr-old, disk.  The simulation by Robertson et al.,
on the other hand, produces a disk galaxy dominated by $\sim$10 Gyr-old
solar-metallicity stars, with little or no spheroidal component.  Clearly, 
measurements of the star formation histories of spiral disks and bulges are
needed to guide the physical prescriptions used in such simulations.

M31, as the nearest massive spiral to our own, offers a unique
opportunity to compare a galaxy's disk and bulge stellar populations;
it is distant enough to ignore its three-dimensional depth, yet nearby
enough to be usefully resolved into stars.  The halo and outer disk of
M31 have been very extensively studied (e.g. Mould \& Kristian 1986,
Hodge et al.\ 1988, Pritchet \& van den Bergh 1988, Davidge
1993, Williams 2002), with recent results providing a mix of main sequence turnoff ages (Brown et al. 2003), a mix
of metallicities (Ferguson \& Johnson 2001, Sarajedini \& Van Duyne
2001, Durrell et al.\ 2004, Rich et al.\ 2004, Sarajedini \& Jablonka 2005), and strong
substructure (Ferguson et al.\ 2002).  The crowding and differential
reddening present in M31's high surface brightness bulge and inner
disk, however, make them very difficult to study (Rich et al.\ 1993, DePoy et al.\ 1993, Rich \& Mighell 1995, Renzini 1998,
Jablonka et al.\ 1999, Davidge 2001, Stephens et al. 2003, hereafter S03), such that
attempts to characterize their stellar populations have met with less
success.  

The only way to overcome the problem of crowding is to observe with
higher spatial resolution, while the effects of differential reddening
are best reduced by observing at longer wavelengths.  Thus, large
ground-based telescopes equipped with adaptive optics (AO) systems and
operating in the near-infrared stand to produce the deepest
observations possible of the extremely crowded inner regions of nearby
galaxies, relatively free of the effects of patchy dust.  In this paper,
we present a derivation of the star formation history
of the bulge and disk of M31 from observations taken with the Altair
AO system and NIRI imager on Gemini North (Davidge et al. 2005,
hereafter Paper I) and by HST/NICMOS (S03).  While in Paper I and S03 we outlined the broad parameters of M31's bulge
and inner disk stellar populations, in this paper we perform an
automated fit of a set of model isochrones to the data to objectively
measure the star formation histories as a function of radius in M31.
We describe our observations and data reduction in \S2 and the
intricacies of the NIRI/Altair photometry in \S3.  \S4 discusses our
completeness and photometric errors, while in \S5 we describe the
derivation of the star formation histories and report our results.
Our conclusions are summarized in \S6.

\section{OBSERVATIONS AND DATA REDUCTION}
As described above, we used data from two sources for the analyses in this paper.  First,
we obtained data of three fields located within 9\arcmin ~of the center of M31
and one more distant field
with the Altair AO system and NIRI
infrared camera on Gemini North on the nights of 18-19 November 2003,
as part of Gemini's System Verification program.  These data, the
details of which we presented in 
Paper I, yielded
$JHK$ images with resolutions of 0\farcs16 ($J$), 0\farcs10 ($H$) and 0\farcs09 ($K$) at
radii within 9\arcsec ~of the central guide stars, becoming worse
towards the edges of the 22\farcs5$\times$22\farcs5 field.  The images
are not quite diffraction-limited, as they were taken before the
source of an internal vibration in Altair was discovered; NIRI/Altair is
now capable of producing images with slightly higher resolution
than we achieved with our data.  For the purposes of this paper, we
used only the $H$ and $K$ images, which are considerably deeper than
the $J$ images, and did not consider the distant Disk 1 field, which does
not have enough sources in it to warrant more analysis than that
already presented in Paper I.  The second source of data used here are the $J$ and $K$ photometry
derived by S03 from {\it HST/NICMOS} observations of 9 fields in M31.  Figure \ref{image}
shows the locations of the NIRI/Altair and NICMOS fields with respect to M31.  Example images of the NIRI/Altair fields can be seen in Figs. 1 and 2 of Paper 1.

We reduced the
NIRI/Altair data independently of Paper I, but following the basic steps outlined
therein.  In brief, we followed the standard NIRI imaging reduction
example found in the Gemini IRAF package to remove the dark current,
create and subtract a sky image, create and divide by a flat field,
correct for bad pixels, and align and combine the individual exposures
into single images.  The step that required the greatest care was the
stacking of the individual dithered exposures, as NIRI/Altair images contain 
some optical distortion.  We corrected for this by allowing the rotation and linear scale of individual images to vary slightly during stacking; the corrections typically amounted to $\sim$1 pixel over the entire field of view.
Once aligned, the
images were averaged together with no scaling applied; as all of the M31
images were taken at airmasses between 1.1 and 1.2, the difference
in atmospheric extinction between images should be $<$1\%.  

\section{PHOTOMETRY OF NIRI/ALTAIR IMAGES}
We made photometric measurements on the combined NIRI/Altair M31 images
using DAOPHOT (Stetson 1987) and ALLSTAR (Stetson \& Harris 1988).
We identified stars using only one pass through the FIND procedure.
We constructed PSFs using $\sim$200 bright stars in each image, using
an iterative procedure so as to remove contaminating fainter
companions.  The PSF was allowed to vary
quadratically with position.  We defined the PSFs out to radii of
0\farcs22 (10 pixels), or $\sim3\times$ the FWHM.  Because of the
incomplete correction for atmospheric turbulence provided by AO, a
significant fraction of the light in the PSF is distributed over the
scale of the seeing disk, which has a FWHM of $\sim$30-40 pixels (0\farcs66 - 0\farcs88) in
our images; the Strehl ratio, in other words, is $<$1.  Moreover, 
anisoplanaticity causes the fraction of light
scattered into this larger halo to increase with increasing distance
from the guide star.  Because our M31 images are extremely crowded, we
cannot correct for the entire contributions of the halos using the
images themselves.  Instead, we applied a partial correction to our
PSF photometry by measuring aperture photometry of the bright PSF
stars out to a 15-pixel (0\farcs33) radius after neighboring stars had been
PSF-subtracted.  Figure \ref{apcor} shows an example of the differences between
the PSF magnitudes and the aperture magnitudes, which are a clear
function of radius.  We fit least-squares linear relationships to the
PSF minus aperture magnitude differences for each of the images and applied
them to the PSF photometry.  In order to keep the corrections small,
we excluded photometry for any stars outside 400 pixels (8\farcs8) in radius from
the central guide stars; we also excluded photometry of stars closer than 20
pixels (0\farcs44) to the guide stars, as these are contaminated by the guide star halos.

Next, we calibrated the photometry by two independent methods.  The
first method was to use aperture photometry of the photometric standard stars
FS103 (Hawarden et al.\ 2001) and AS40-1 (Hunt et al.\ 1998), which were observed on the same nights
as the M31 fields and at similar airmasses, for a total of 12 independent observations in each filter.  From the uncrowded
standard star images, we found that an aperture of 100 pixels (2\farcs2) in
radius was needed to enclose $>$99\% of the light from the stars (as judged from the asymptotic value of the growth curve), with
standard deviations of 
0.02 ($H$) and 0.03 ($K$)
magnitudes in the zero point using this aperture.  Using the smaller
15-pixel radius aperture, we found that we needed to apply corrections
of 
0.40$\pm$0.09 ($H$) and 0.32$\pm$0.06 ($K$)
magnitudes to account for the light outside the aperture.  The larger
standard deviations of the photometry with the smaller apertures are
the result of temporal Strehl ratio variations among the set of
standard star images, and reflect the true uncertainty of our
photometric zero points introduced by changing atmospheric conditions during our observations.

The second calibration method was to fit our PSFs to the wings of the saturated
central guide stars, which have accurate photometry from 2MASS (Skrutskie et al.\ 1997),
and are several magnitudes brighter than the M31 background, making crowding a
non-issue.  The advantage of this approach is that it avoids the
uncertainty of tying the calibration to observations taken at a
different time and under different conditions.  The disadvantages are
that it relies on a large extrapolation of the contribution of the
saturated portion of the guide stars (albeit an extrapolation based on
information from 200 stars in each image), and that the guide stars, while
bright, are not well-established standards.  Despite these drawbacks,
we felt that this second approach provides a useful check on the
reliability of our calibration.  As shown in Table 1, the two
calibration methods agree well to within the uncertainties, the
exception being the $K$ zero point of the Disk 2 field, where they
differ by 0.25 magnitudes.  As the guide star in this field is the
brightest, with 55\% of its light in the saturated core, we attribute
the difference either to an error in the PSF extrapolation or, as
suggested in Paper I, to intrinsic variability of the guide star
itself.  We thus rely on the standard star calibration for the remainder of
this paper, and note that the independent check of using the guide stars to calibrate the photometry gives us confidence that our zero points are accurate to at least 0.1 magnitude, if not better.

Once the photometry was calibrated, we produced color-magnitude
diagrams (CMDs) by finding matches within $\sim$2 pixels of each other from
the lists of $H$ and $K$ photometry.  Figure \ref{cmds} shows the CMDs.  The
zero points of these CMDs agree to within $\sim$0.1 magnitudes with those published in
Paper I; the main difference between them is the use of an aperture
correction that varies as a function of position in the current
paper.  
We can check whether the photometry published
here represents an improvement over Paper I by plotting $H-K$ vs.\ radius, as is
done in Figure \ref{hkr}; because the red giant branch (RGB) and asymptotic giant branch (AGB) sequences are nearly
vertical in the $H-K,K$ plane, and we do not expect population gradients
across the small NIRI/Altair field of view ($\sim$85 pc at the
distance of M31), residual Strehl variations will likely appear as
gradients in $H-K$.  Figure \ref{hkr} shows that small $H-K$ gradients exist
in the photometry published in Paper I, but have been successfully
removed in the photometry presented here.  

\section{COMPLETENESS AND ERROR ANALYSIS OF NIRI/ALTAIR PHOTOMETRY}
We calculated the completeness of our photometry and the photometric
errors through two sets of artificial star simulations.  The first
employed the ``traditional'' technique of adding a sample of artificial stars to multiple copies of the original $H$ and $K$ images and
rerunning the steps of our photometric pipeline on each of these
images.  We selected the input stars by choosing 20000 stars over the luminosity range $-7.3\le M_K \le 1.0$ from a 10-Gyr
solar metallicity Girardi et al.\ (2002; hereafter G02) isochrone, which we distributed 50 at a time to 400 copies of the original $H$ and $K$ images.  
We did not consider inserting stars with a range of color, as the photometric errors and completeness are almost entirely functions of magnitude.
We converted the absolute $H$ and $K$ magnitudes to apparent
magnitudes assuming an M31 distance modulus of 24.45 (e.g. Freedman \& Madore 1990, Stanek \& Garnavich 1998, Mochejska et al.\ 2000), applied our
photometric calibration in reverse to bring the stars onto the
raw magnitude system, and assigned random positions to the stars.
After using ADDSTAR to place the artificial stars in the images
and running our DAOPHOT/ALLSTAR pipeline on the artificial images, 
the output lists of photometry were culled to exclude matches with
stars with positions within 0.5 pixels and with magnitudes within 0.1 of
the original photometry.
Artificial stars were considered recovered if their entry in the input
list contained a match in the culled output list
within 2.5 pixels (0\farcs055) and contained a
counterpart in the other filter within $\sim$2 pixels, the same matching
radius used to create the observed CMDs.  These radii were set through an examination of of the distributions of the position differences of matches.

The second set of simulations consisted of photometry of images
containing entirely artificial stellar populations, as done e.g.\ by
S03, and designed to match our real images in field of view and surface brightness.  
For these simulations, we selected stars from a 12-Gyr solar
metallicity G02 isochrone with a Salpeter (1955)
IMF, and included stars as
faint as $M_K=4$.  These faint stars lie roughly 7.5 magnitudes below
our detection limit, but were included to accurately model the fluctuations in 
background.  We produced simulated $H$ and $K$ images for each of the
Bulge 1, Bulge 2, and Disk 2 fields.  We selected a
sufficient number of stars
to match the total luminosities derived from Kent's (1989) $r$-band
surface brightness model,
assuming $r-K=2.9$, $(m-M)_0=24.45$, and $A_K=0.024$; for
the Bulge 1 field, 6.6$\times10^6$ stars were simulated, while the Bulge 2 and
Disk 2 fields contain 2.8$\times10^6$ and 1.1$\times10^6$ stars, respectively.  Our choice of an old isochrone ensures that our simulations represent the most severe possible crowding; had we used a younger isochrone, the number of stars needed to match the observed surface brightnesses would be smaller, and the crowding less severe.
We assigned
uniformly distributed
random positions to the stars, which is appropriate because the
NIRI/Altair field of view is small compared to M31's surface brightness
gradient.  We did not use ADDSTAR to insert the stars into images, as
this would have been too time-consuming.  Instead, we produced 16 different versions of each of the artificial images, each one convolved with the DAOPHOT PSF calculated at the different positions of a 4$\times$4 grid with grid spacing of 4\farcs4 (200 pixels).  We then combined the 16 different versions together using a weighted average that resulted in an image with the PSF bilinearly interpolated over the 4$\times$4 grid, thus approximating the PSF variability in our real images.
After adding Poisson noise from the stars and sky, photometry of these images was produced using the same DAOPHOT/ALLSTAR
pipeline used for the real M31 images and the same photometric
calibration; the CMDs were produced by the same
process described in \S3.  For the purposes of computing
completeness and photometric errors from the simulation, we paired each star
measured by ALLSTAR with
the brightest star in the input list found within 2.5 pixels (0\farcs055).  We picked the brightest star rather than simply the closest star, as within the area of a resolution element, the brightest star contributes the majority of the flux to the output image.

Figure \ref{simcmds} shows the similarity of the depth of
the artificial CMDs to the real CMDs, to be compared to Figure \ref{cmds}.  The scatter at faint magnitudes is also reproduced by the simulation; for example, at $K\sim19.5$, $\sigma_{H-K}$ is 0.14 in the observed Bulge 1 CMD and 0.12 in the simulated one.  At bright magnitudes, the observed widths of the RGB and AGB branches are much wider than expected by purely photometric errors; for example, at $K\sim17.5$, $\sigma_{H-K}$ is 0.074 in the observed Bulge 1 CMD and 0.024 in the simulated one.  This could indicate metallicity dispersion and/or differential reddening, as noted in Paper I.
Figure \ref{errfig} shows the average photometric errors and completeness derived from
the traditional artificial star tests; those derived from the entirely simulated images are very similar, and so are not plotted.  We
also show the photometric errors predicted by the Olsen et al.\ (2003) analytical model of crowding.  The good agreement
between the analytical model and the simulations at the bright end
demonstrates that crowding dominates the photometric errors at bright
magnitudes.  The model correctly predicts that the errors in $H-K$
color are {\em smaller} than those in $K$ magnitude, as only happens
in the case of crowding, where photometric errors in different filters are
correlated.  As concluded in Paper I, the crowding is not severe enough to 
explain the large number of bright red stars in the M31 CMDs as being
photometric blends.  Very few such stars are seen in the simulated
CMDs in Figure \ref{simcmds}, as is expected because the crowding-induced errors
are $\lesssim$10\% down to $K\sim$20.  At fainter magnitudes, the
predictions of the analytical crowding model fall below the errors in
the simulation, indicating the increasing contribution of Poisson
noise to the errors.

\subsection{Comparison of NIRI/Altair and NICMOS Photometry}
In Figure \ref{nicvsniri}, we show the combined $(H-K)_0, M_K$ CMD of the 9 NICMOS fields from S03 compared to that of the three NIRI/Altair fields presented here.  In constructing the CMDs, we have employed a distance modulus $(m-M)_0=24.45$, dust extinction $A_K=0.1$, and $E(H-K)/A_K=0.571$.  The excellent match between the NICMOS and NIRI/Altair CMDs gives us confidence that our $H$ and $K$ photometry is well-calibrated.  Plotting a histogram of color for all magnitudes $-6.25 \le M_K \le -5.75$, we find that the two distributions have averages of $<(H-K)_{0,{\rm NICMOS}}>=0.322$ and $<(H-K)_{0,{\rm NIRI/Altair}}>=0.285$, with standard deviations of 0.064 and 0.081 magnitudes, respectively (discounting points further than 3$\sigma$ from the means).  

We are concerned, however, about the $J$ photometry published in S03.  Figure \ref{nicvslmchk} compares the combined M31 NICMOS $(H-K)_0, M_K$ CMD with the $(H-K)_0, M_K$ CMD of the LMC Bar from the 2MASS Point Source Catalog (Cutri et al.\ 2003), where we have used $(m-M)_{0}=18.5$ and $A_K=0.025$ (Zaritsky et al.\ 2004) for the LMC Bar.  The histograms of color for stars with $-6.25 \le M_K \le -5.75$ clearly indicate that the LMC sequence lies $\sim$0.1 magnitude blueward of the M31 sequence.  If we ascribe the difference in color to lower metallicity and/or younger age in the LMC, then we expect the $(J-K)_0, M_K$ LMC CMD to be $\sim$0.2 magnitudes bluer than that of M31.  However, as seen in Figure \ref{nicvslmcjk}, the RGB and AGB stars in the LMC are actually $\sim0.05$ magnitudes {\em redder} in $(J-K)_0$ than in M31.  We thus suspect that the $J$ photometry of S03 is too bright by as much as 0.25 magnitudes, and thus do not consider it in this paper.  This suspicion is deepened by the fact that the G02 isochrones correctly predict the $(J-K)_0$ fiducial colors of globular cluster sequences from Ferraro et al.\ (2000) to within 0.03 magnitudes, ruling out error in the isochrone colors.

\section{STAR FORMATION HISTORIES}

Over the past few years, extracting star formation histories from observations of mixed stellar populations, a task that Schwarzschild (1965)
described as ``somewhat hopeless", has become routine (see
Dolphin 2002 for an overview).  There are many reasons for this
change in outlook.  Perhaps most important is the availability of digital detectors yielding excellent semi-automated photometric measurements; also important are
the increased resolution of modern telescopes, better understanding of
stellar structure and evolution, and the recent development, coupled
with greater computing power, of automated techniques that measure the
star formation histories by statistical comparisons of models and
observations (e.g. Tolstoy \& Saha 1996, Aparicio et al.\ 1996, Dolphin 1997, Hernandez et al.\ 1999, Harris \& Zaritsky 2001).  While these techniques differ in detail, all
rely on the ability to generate synthetic observations from model isochrones,
given a set of parameters (e.g. age, metallicity, distance, reddening,
binary properties, and IMF slope) describing the model stellar
population.  An important step in the production of the synthetic observations
is the modeling of the photometric errors and completeness; as seen
above, these are influenced by a combination of crowding and
instrumental noise, and are best estimated through artificial star
tests.  Finally, all of the techniques derive the star formation
history by computing the likelihood that the observations are a
representation of a given model.  Where the techniques differ most is
in their approach to searching the likelihood space for the best
solution.  

In this paper, we derive the first detailed star formation
histories of the bulge and inner disk of M31.  To this end, we used
the software developed by us in Olsen (1999) and Blum et al. (2003)
(which is based on the technique outlined by Dolphin 1997 and Dolphin
2002) to fit model stellar populations to the $K$ luminosity
functions (LFs) of our fields.  We assumed that our
LFs may be modeled by linear combinations of G02
isochrones specified by discrete ages from the set \{1, 3, 5, 10\} Gyr
and discrete metallicities from the set Z=\{0.0001, 0.0004, 0.001,
0.008, 0.019, 0.03\}.  Our choice of basis isochrones and our choice
to model the $K$ LFs instead of the complete CMDs were driven by
experiments, which will be described fully in a forthcoming paper, in
which we used our software to derive the star formation history of the
Large Magellanic Cloud (LMC) from 2MASS (Cutri et al.\ 2003) photometry of its
evolved stars.  In brief, we found that by using only the $K_s$ LF and the G02 isochrones, our results
were qualitatively consistent with star formation histories
derived from deep optical {\it HST} photometry; fits to the full 2MASS CMD of the LMC did not provide as good agreement.  This is in part because the colors in the models adopted here employ solar-scaled opacities, which are inappropriate for chemically processed AGB stars (Marigo et al.\ 2003).  These results are in general agreement with Cioni et al.\ (2005), who performed similar analysis of the 2MASS LMC photometry; Marigo et al.\ (2003) show the effort needed to qualitatively reproduce the features in the full 2MASS LMC CMDs.

We employed a
distance modulus of $(m-M)_0=24.45$ and a Salpeter (1955) IMF throughout our analysis.  We
set $A_K=0.1$ on the basis of work by Xu \& Helou (1996) and Haas et
al. (1998), who measured the optical depth due to dust in M31 through
modelling of the emission detected by IRAS and ISO.  Haas et al.\ (1998) found an average $\tau_V\approx0.5$ over most of the M31 disk, with higher $\tau_V$ in the 10 kpc ring and in a knot 5\arcmin ~from the nucleus; none of our fields, however, lie within this knot, and only F2 and F280 (S03) lie near the 10 kpc ring.  Xu \& Helou (1996) found average $0.7\lesssim\tau_V\lesssim1.7$ over the range of radii of our fields, with $\tau_V\sim0.8$ being appropriate for most of our fields.  Taking $A_V=1.086\tau_V$ and adopting $A_K/A_V=0.11$ (Schlegel et al.\ 1998), we find $A_K\approx0.1$.  We note, however, that this represents only an average, whereas the true extinction could explore values between $0.05\lesssim A_K\lesssim0.2$.
The Milky Way foreground extinction of $A_V=0.062$ (Schlegel et al.\ 1998) makes a negligible contribution to the total absorption, and so is neglected here. 
For these data, we did
not bother including different IMF slopes or binary star
fractions in our models. Because the mass range spanned by the upper
RGB and AGB is very small, changing the IMF would affect only the
total mass represented by the star formation history.  In addition, the
contributions of binary companions to the luminosity of bright RGB and
AGB primaries, which are in a short-lived phase of stellar evolution,
are negligible.

Given the above assumptions, our goal was to find the combination of
age and metallicity with maximum likelihood of
representing our observations.  To do this, we first computed the
theoretical $K$ LFs corresponding to each combination
of model parameters.  The first step
in this computation was to produce separate $K$ LFs for each of
the 24 combinations of age and metallicity from the G02 isochrones; in
doing this, we assumed a width of 1 Myr for each
isochrone (yielding a practically instantaneous burst) and a star formation rate of 1 M$_\odot$ yr$^{-1}$.  Using a wider width in age for the isochrones would not affect our results, as the $K$ LF has only coarse sensitivity to age and metallicity.
We used a bin size of 0.16 in $K$, which is small enough to resolve the
differences between the input isochrones.  Next, we translated each of
the LFs to the distance of M31 and added the effect of extinction by dust.  We
then convolved each of the LFs with an adaptive Gaussian
kernel representing the photometric errors derived from our artificial
star tests, and multiplied them by a curve representing the
completeness of our observations as a function of observed $K$.  Finally, we trimmed the model LFs to include only the
portions with completeness greater than 50\%; doing this ensures that our results are not adversely affected by incompleteness and photometric bias.  These 50\% limits are listed in Table 2; with the exception of F1, these limits include RGB tip stars of all ages and metallicities considered here.

Figure \ref{modlf} illustrates the population-sensitive features of the $K$ LF.  At higher metallicity, the RGB tip becomes considerably brighter in $K$, while the AGB tip brightens also.  The RGB tip becomes brighter with increasing age for ages below $\sim$5 Gyr, while the AGB tip becomes fainter with increasing age at all ages.  Because none of these trends are linear, there is hope of extracting both age and metallicity from the detailed shape of the $K$ LF.

After binning and trimming our observed $K$ photometry in the same way as was
done for the model LFs, we used the minimization routine
{\tt amoeba} (Press et al.\ 1992) to find the model with maximum likelihood of
representing our observations.  We computed the likelihood assuming
Poisson-distributed data, as discussed by Dolphin (2002); in practice,
this meant minimizing the parameter $\chi_\lambda^2 = 2\Sigma_i m_i -
n_i + n_i ln(n_i/m_i)$ (Mighell 1999, Dolphin 2002), where $m_i$ is the number of
stars predicted by the model in the $i$th bin and $n_i$ is the number
of observed stars in the bin.  We
allowed the routine to form a linear combination from any of the 24
models representing different ages and metallicities.
The coefficients of this linear combination are the best-fit
star formation and chemical enrichment history, given our assumptions about $A_K$, the distance modulus, and the IMF.
We also computed fits adopting $A_K=0.0$ and $A_K=0.2$, so as to explore the effect of possible errors in our assumptions about the dust extinction and/or zero point errors in our $K$ photometry.
In order to verify that having both age and metallicity as free parameters is warranted, we also computed fits to
the LFs in which only one of the two variables (age, metallicity) was
allowed to vary, the other one being held at a constant value.

For the solutions with $A_K=0.1$ and having full freedom in age and
metallicity, we derived errors in the fitted star formation rates by
running our fitting procedure on bootstrapped samples of our original
data.  These samples were created by randomly drawing $N$ stars from
our original CMDs, where $N$ is the number of stars in the original
CMD, but permitting repeat draws of the same stars.  From the best-fit
star formation rates for these bootstrapped samples, we calculated the
1-$\sigma$ standard deviations of the star formation rates in each
bin; we call these our measurement errors.  A separate question is
whether our best-fit models are statistically likely representations
of our observations.  To address this question, we computed the expectation value and variance of the $\chi_\lambda^2$ parameter from our best-fit models, and then calculated the probability $P_\lambda$ that values of $\chi_\lambda^2$ as high as the ones we measured could arise by chance.  As discussed by Dolphin (2002), in the limit of large $n_i$ and $m_i$ (in which a Poisson distribution is indistinguishable from a Gaussian), $\chi_\lambda^2$ has an expectation value of 1 and a variance of 2 in each bin $i$; for smaller $m_i$, the expectation value and variance deviate slightly from these values.

Table 3 summarizes our fits of the population mixes in the Bulge 1,
Bulge 2, and Disk 2 fields observed with NIRI/Altair and the NICMOS
fields from S03.  The pieces of information included in the table are
an identifier for each solution, the $A_K$ used in the solutions, the
best-fit star formation rates in each of the age and metallicity
components with errors in parentheses where calculated, the
$\chi_\lambda^2$ for each
fit, a number $\sigma_{\chi_\lambda^2}$ representing the extent to which the fitted $\chi_lambda^2$ deviate from the expectation values for the models (calculated as $(\chi_\lambda^2-<\chi_\lambda^2>)/<\sigma>$, where the bracketed quantities are the expected mean and variance of $\chi_\lambda^2$ for each model), and the probability $P_\lambda$ (in \%) that $\chi_\lambda^2$
as high as the ones listed could have arisen by chance.  If the LFs were truly drawn from the given models, then we expect $P_\lambda\sim$50\%; low values of $P_\lambda$ are indicative that the models are not good fits.
The
identifiers consist of a combination of the field name and a number
between 1 and 13.  The number 1 is reserved for the solutions with
$A_K=0.1$ and full freedom in fitting the ages and metallicities;
numbers 2 and 3 solutions with $A_K=0.0$ and 0.2, respectively;
numbers 4-7 solutions with $A_K$=0.1 and ages held to fixed values;
and numbers 8-13 solutions with $A_K$=0.1 and metallicities held to
fixed values.  

Our preferred solutions are those with $A_K=0.1$, as we
believe this to be closest to the correct value of the dust
extinction.  However, in many cases one or the
other of the $A_K=0.0$ and $A_K=0.2$ solutions provide better
statistical fits.  We examine the effect of adopting these other fits in \S5.2
There are also four fields, F2, F3, and F4, F280, where
one or more of the fixed-age/fixed-metallicity solutions provide
nearly as good fits.  However, at least in the cases of F3 and
F4, these other solutions are not qualititatively different from the
solutions with full freedom in fitting the ages and metallicities.
Thus, on the whole, there is good evidence that, given our assumption about $A_K$ and the distance modulus of M31, a mix of ages and metallicities are required to model its stellar populations.  In the case of F280, however, the statistical likelihood $P_\lambda$ of many of the fits are so high that we can only conclude that F280 is not well fit by exclusively 1 Gyr, 5 Gyr, or 10 Gyr-old populations.

Finally, in seven of the fields (Bulge 1, Bulge 2, F1, F4, F170, F174, and F177) the statistical likelihood $P_\lambda$ of the fits is $<$1\%, with F4 and F174 being the worst.  In these cases, we found that we could typically obtain an order of magnitude higher likelihood if we excluded the faintest 1-2 bins of the LFs, or, in the case of F4, the bins containing the brightest two stars.  Running new fits with these bins excluded did not appreciably change the best-fit star formation histories, so for overall consistency, we report only the fits to the full LFs.  We suspect, however, that the reason for the low $P_\lambda$ of some of our fits is error in our completeness and photometric error measurements at faint magnitudes.

\subsection{Age and Metallicity Resolution}

To demonstrate the age and metallicity resolution of our data, 
we performed three sets of tests using simulated observations
drawn from the set of G02 isochrones.  Our first
test consisted of a model population with uniform extinction
$A_K$=0.024, four bursts of star formation at 10, 5, 3, and 1 Gyr, and
increasing metallicity with age, and is depicted as a ``population
box'' (Hodge 1989) in Figure \ref{test1}.  Stars were drawn randomly from this
population in numbers to match the surface brightness of our Bulge 2
field; these stars were then paired with stars from our artificial
star tests, the results of which we used to apply photometric errors
and to decide which of the simulated stars would be lost through
incompleteness.  The result was an LF with properties similar to that
of our Bulge 2 field, but with a known input population mix.  
Figure \ref{test1} shows the recovered population box obtained by running the artificial LF through our software,
 and compares the recovered age and metallicity
distributions with the input ones.  We see that our solution closely
recovers the overall age and metallicity distributions of the input
population.  In addition, with the exception of the oldest age bin,
the dispersion in metallicity at a given age is of order our bin size.  

Our second test isolated the oldest populations, which are the most
difficult to discriminate.  Our input population, as shown in Figure
\ref{test2}, consisted of bursts at 5 and 10 Gyr and solar metallicity.  
The fit recovered approximately the
correct age and metallicity distributions, with a dispersion of
order the bin size in the fitted metallicities, and demonstrates that we
can faithfully discriminate between intermediate-age (5 Gyr) and old
(10 Gyr) populations.

Up to this point, the tests we have described employed exactly the
same isochrone set to produce the input populations as were used to
derive the solutions, whereas nature is unlikely to be so cooperative.
To test the effect of this idealization, we conducted a third test
where the input population approximated a constant star formation rate
from 1 to 10 Gyr with linearly increasing metallicity.  We used the
G02 isochrones, interpolated onto a grid with 0.1-Gyr
spacing, to produce the input population, while for our solution, we
continued to use only the isochrones from the discrete sets age=\{1,
3, 5, 10\} Gyr and Z=\{0.0001, 0.0004, 0.001, 0.008, 0.019, 0.03\}.
In Figure \ref{test3}, we show both the
input population box and that recovered by the best-fit solution to
the simulated $K$ LF.  With the exception of the addition of
a range of metallicities in the 10 Gyr age bin, our solution recovered
the approximate metallicity distribution; it also recovered roughly the
correct star formation rates in the 1-5 Gyr range.  However, the star
formation in the 10 Gyr bin far exceeds the input value; indeed, it
appears that all of the star formation with ages 6 Gyr and above was
drawn into the 10 Gyr bin.  This test thus shows that while we can
pick out $\sim$5 Gyr populations with our basis set of models,
populations older than $\sim$6 Gyr will all appear to have the same
age.

\subsection{Results and Discussion}

As an illustration of the results, Figure \ref{popbox} shows the
population boxes (Hodge 1989) of the best fits to the NIRI/Altair
observations and to one of the NICMOS fields.  We also show the model
LFs compared to the observations as well as a comparison of the model
and observed CMDs, expressed as Hess diagrams, where we have assumed
$E(H-K)=0.571A_K$ (Schlegel et al.\ 1998).  As seen in the figure,
while the models provide good fits to the LFs, the models we used are
not designed to fit the full near-infrared CMDs; the data consistently
fall to the red of the models by $\sim$0.04 magnitudes in the mean
color.  This color difference is small enough to be unimportant for
our conclusions if the error were all in $K$, but would introduce
population differences if we fit the full CMDs rather than the $M_K$
LFs.  We suspect, however, that the problem lies mainly in the model
$H$ magnitudes, because the Ferraro et al.\ (2000) $(J-K)_0$ fiducial
colors of Galactic globular clusters agree nicely with the Padova
isochrones, while the $(J-H)_0$ fiducial colors from Valenti et al.\
(2004) are systematically bluer by $\sim$0.04 magnitudes than the
isochrones.  See Marigo et al.\ (2003) for a deeper investigation of
the physics needed to fit near-IR photometry of evolved stars.

We examine the age and metallicity trends more broadly in Figure
\ref{radplots}, where we plot the mean ages and metallicities (where
we define [M/H]$\equiv\log Z/Z_\odot$) of our fits with $A_K=0.1$
versus radius on the sky/bulge-to-disk ratio as well as the relative
contributions of each of the stellar population components.  Because
our fields contain contributions from both bulge and disk components,
we use radius on the sky rather than deprojected radius in the M31
disk plane for these plots.  We find that the average age of the fits
is $\sim8$ Gyr with an r.m.s. of $\sim$1.5 Gyr at all radii.
Examining the contribution of the individual age components of our
fits, the $\sim$10-Gyr population dominates the fits at all radii and
bulge-to-disk ratios, contributing $\gtrsim$60\% of the star formation
over M31's lifetime in all fields.  Where the 10-Gyr contribution is
lowest, the 5 Gyr population contributes correspondingly more to the
overall star formation, while the 1 and 3 Gyr populations never
contribute more than $\sim$20\% of the total star formation.
Examining the mean metallicities, we find that at a radius of 5\arcmin
~from the center, the mean metallicities are near solar with an
r.m.s. of $\sim$0.5 dex.  The fits suggest that the mean metallicity
drops by $\sim$0.5 dex in fields with B/D $\sim$0.3 -- 1, with an
additional hint that the metallicity is also slightly lower at the
smallest radii.
In the inner bulge, however, this drop in mean
metallicity is due entirely to the sudden contribution of a
substantial [M/H]=-2.3 population, which we believe to be spurious, as
discussed further below.  Examining the individual metallicity
components shows that while all contribute to some degree, the fits
are indeed dominated by the components with [M/H]$\ge-0.5$ at nearly
all radii. Our results are independent of whether the photometry
derives from NIRI/Altair or HST/NICMOS.

We demonstrate the effect of adopting either the fits with $A_K=0.0$ or $A_K=0.2$ in Figure \ref{radplots02}.  The mean metallicities are $\sim$0.25 dex lower in the $A_K=0.2$ fits and $\sim$0.25 dex higher in the $A_K=0.0$ fits than in the case of $A_K=0.1$.  The mean ages are $\sim$1 Gyr younger in the case of $A_K=0.2$ compared to $A_K=0.1$, while the $A_K=0.0$ mean ages are similar, with the exception of F280, the field most distant from the M31 center.  However, as stated previously, this field has a very uncertain star formation history on account of its low number of stars.  In Figure \ref{radplots_best} we show the result of picking those fits with highest $P_\lambda$ and $A_K$ in the range 0.0 -- 0.2.  The mean ages and metallicities shown are nearly identical to those of Figure \ref{radplots}.  We thus conclude that aside from F280, our results are insensitive to changes in $A_K$ at the level of $\pm$0.1 magnitudes.  The conclusion that older ($\sim$6+ Gyr) populations with [M/H]$\sim$-0.5 -- 0.0 dominate M31 for fields with B/D$\ge$0.3 appears robust.

How well does our method break the age-metallicity degeneracy often
found in stellar populations studies?  The tests described in \S5.1 do
not reveal any such degeneracy in the models at our age and
metallicity resolution.  We caution, however, that our breaking of the
age-metallicity degeneracy in some cases depends on the existence or
not of a few ($\sim$10) bright AGB stars; indeed, these are the stars
from which much of our age sensitivity derives.  Also, because the
models are not perfect descriptions of our observations, there exists the
possibility that some of our models are nearly degenerate with respect
to the data or that models that are not
representative of the true populations could drive the
fits.  In these cases, we would expect to see correlations between
pairs of model parameters fit to different fields.  The right-hand
panels of Figure \ref{radplots} show that there does appear to be
correlation between some of the components in our population fits,
e.g. between the 5 and 10-Gyr components in the upper right-hand panel
of the Figure.  To quantify the degree of correlation, we calculated
the Spearman correlation coefficients $r_s$ between all pairs of
population components for our 12 fields.  The $r_s$ values are listed
as a matrix in Table 4a, while Table 4b lists the probability that
correlations as strong as indicated by each $r_s$ value could have
occurred by chance.  Ignoring the trivial diagonal elements of the
matrix, we see that there are a few components that have strong
anticorrelations with $<$10\% probability of being chance
correlations: the 5- and 10-Gyr components, the 10-Gyr and Z=0.03
components, the 1-Gyr and Z=0.008 component, the Z=0.0004 and Z=0.019
components, and the Z=0.019 and Z=0.03 components.  These component
pairs are thus probably at least partially degenerate; however, these
anticorrelations do not explain away the dominance of 6+ Gyr,
solar-metallicity populations in the bulge and inner disk.  We also
find a significant positive correlation between the 1-3 Gyr and
Z=0.0001 components.  This correlation suggests that the young and
intermediate-age (1-3 Gyr), very metal-poor ([M/H]$\sim -2$)
populations that appear in some of the fits are due to an incomplete
description of the data by our model set.

\subsection{Comparison with Other Work}

As demonstrated by DePoy et al. (1993), Renzini (1998), and S03,
seeing-limited ground-based observations of the bulge and inner disk
of M31 (e.g. Rich et al.\ 1993) do not have sufficient
resolution to decipher the full stellar population mix.  The recent
work by Sarajedini \& Jablonka (2005), who derived M31's bulge metallicity
distribution from $V$ and $I$ observations taken with HST/WFPC2, thus
provides the most direct comparison with our results.  
From the $V-I,V$ CMD of a field near F170, Sarajedini \&
Jablonka found a metallicity distribution peaked at [M/H]=0.0, with a
gradual tail to [M/H]=$-1.0$ and a more rapid decline at higher
metallicities.  In Fig.\ \ref{sfh170}, we show the metallicity distribution function and population box we derived for F170.  Although our metallicity resolution is coarser than theirs, the distribution functions are in qualitative agreement.  We do not find that all of the populations have ages as old as 10 Gyr, as assumed by Sarajedini \& Jablonka, although these old populations do comprise the biggest component.

In Fig. \ref{sfrtot}, we show the population box
integrated over all fields from our work; it is dominated by the high
surface brightness bulge.  Ignoring the possibly spurious intermediate-age
metal-poor component, we measure a metallicity distribution function
that is a bit more sharply peaked than Sarajedini \& Jablonka, but is also in
basic qualitative agreement.  It is worthwhile to note that the 3-5 Gyr, [M/H]$\sim -2$ component is strongest in the Bulge 1, F1, and F177 fields, and weak or non-existent in the other fields.  Examining Figure 1, these fields are the three nearest the center of M31 and all lie near each other on the sky.  It is perhaps tempting to speculate that the metal-poor component, if real, represents the remains of a dwarf galaxy that has merged with M31.  However, in the absence of dust, such a component would make the integrated color of M31 $\sim$0.2 magnitudes bluer in $B-V$ than a purely 10 Gyr-old, solar-metallicity population, which seems ruled out by integrated surface photometry (Sandage et al.\ 1969).  More likely is that the component appears when the crowding is most severe and the photometry is shallower.

We can also make indirect comparisons
with other work.  Davidge (1997) performed integrated spectroscopy of
the inner few arcminutes of M31's bulge, and found Balmer indices
consistent with old ($\sim$10 Gyr) stellar populations and Fe and Mg
line strengths indicative of metallicities in excess of solar, in
agreement with the conclusions of Spinrad \& Taylor (1971).  From our
fits, we find a luminosity-weighted age of $>$8 Gyr for our fields
with $B/D>3$.  If we consider the full fits, then the
luminosity-weighted metallicity is [M/H]=$-0.7$; if we exclude the
most metal-poor component, it is [M/H]=$0.0$, in good agreement with
the integrated spectroscopy.  Venn et al. (2000) measured elemental
abundances in four young, massive supergiants from high-dispersion
spectra, finding [M/H]$\sim$0.0.  Jacoby \& Ciardullo (1999) and
Richer et al.\ (1999) derive median [O/H]$\sim-0.6$ in
M31 bulge PNe with a dispersion of $\sim$0.25 dex, which would be in
agreement with the sense of our result if we included the most
metal-poor bulge components.  However, Jacoby \& Ciardullo point to
several factors that could bias the planetary nebula samples toward
lower metallicities.  Williams (2002) measured the star formation
histories from WFPC2 archival images of 27 fields at radii greater
than 20$\arcmin$ from the center of M31.  In his innermost fields,
Williams found star formation histories dominated by old populations
and metallicities typically in the range $-0.5\lesssim$[Fe/H]$\lesssim0.0$, in
agreement with our inner disk and bulge star formation histories.  Thus, on the whole, our derivation of M31's bulge and disk stellar populations--made solely on the basis of ground-based AO and HST/NICMOS $K$ observations--are in excellent agreement with work done at optical wavelengths.

If we compare the star formation histories
and age-metallicity relations that we derive for M31 (Figure
\ref{simcomp}) with the predictions of the simulation by e.g. Robertson et
al.\ (2004), we conclude that the simulations appear to be on the
right track to predicting the properties of galactic disks; however,
we need higher quality data and improved understanding of the late
stages of stellar evolution in order to provide stronger constraints
from observations of nearby galaxies.  We do not yet have the precision necessary to rule out the results of other simulations, such as Abadi et al. (2003).

Comparing our results for the M31 bulge with that of the Galactic
bulge, we find similar stellar populations in both.  Deep imaging
studies of the Galactic bulge indicate that the bulk of the bulge is
relatively old (Zoccali et al. 2003; Feltzing \& Gilmore 2000), with
an age that appears to be comparable with that of metal-rich globular
clusters.  The full Galactic bulge star formation history includes a
fraction of stars formed at intermediate ages (Blum et al.\ 2003, van
Loon et al.\ 2003), as we see in M31, but 75\% have ages older than 5
Gyr.  Our results confirm those of Sarajedini \& Jablonka (2005), who
found remarkably similar M31 and Galactic bulge metallicity
distributions, as long as we ignore the possibly spurious intermediate-age
metal-poor component that appears in our solution.  Comparing the M31
and Milky Way disks, by contrast, we find significant differences.
Although all of our M31 fields contain a mix of some bulge and disk
stars, making a unique decomposition impossible, the lack of a radial
age gradient suggests that the evolution of the star formation rate
with time in the disk has been similar to that of the bulge.  In the
Milky Way, the disk star formation history appears to have been much
more constant (e.g. Hernandez et al.\ 2001).

\section{CONCLUSIONS}
We have discussed $H-K,K$ CMDs of the bulge and disk of M31
obtained with the Altair AO system and NIRI instrument on Gemini
North; these are the deepest CMDs obtained to date of the inner
regions of M31, and demonstrate the promise of ground-based AO for
stellar populations work in the crowded regions of nearby galaxies.
We have combined our $K$ LFs with the {\it HST/NICMOS} $K$ LFs
published by S03 and have derived the coarse star formation history of
M31's bulge and disk. By constraining the extinction to values
consistent with emission by dust in the IRAS and ISO bands, we were for many of our fields
able to produce statistically acceptable fits using a parameter set
that included isochrones with $Z$=\{0.0001, 0.0004, 0.001, 0.008,
0.019, 0.03\} and age=\{1, 3, 5, 10\} Gyr.  These fits revealed a
bulge dominated by old ($\gtrsim$6 Gyr) solar-metallicity stars, in
agreement with past work on the M31 bulge (Spinrad \& Taylor 1971, Davidge 1997, Venn et al.\ 2000, Davidge et al.\ 2005, Sarajedini \& Jablonka 2005).  We find evidence
for $\sim$0.5 dex lower metallicities in the disk,
which is in accord with recent work indicating [Fe/H]=$-0.7$ in the
outer disk and inner halo (Ferguson \& Johnson 2001, Sarajedini \& Van Duyne
2001, Durrell et al.\ 2004, Rich et al.\ 2004).  Finally, we found that old populations, which we
again define as having age $>$6 Gyr, dominate the star formation
histories at all radii.  Thus, we suggest that a lower bound on the redshift of
formation of the disk is $z=0.7$.  Our analysis detects no age
difference between the bulge and disk to the limit of our precision,
leaving open the possibility that the disk formed the bulk of its
stars at still higher $z$.  The same conclusions are reached if we choose to consider the NIRI/Altair and NICMOS results separately.

Although we consider that the overall agreement of our results with that of past work represents a real success for ground-based AO-corrected imaging and photometry, there is plenty of room for improvement.  On the observational side, the effects of anisoplanaticity are strong in our NIRI/Altair images; a wider field of view and more slowly varying PSF would provide definite improvement in the photometry.  The recent addition of a field lens that conjugates Altair to a layer of turbulence nearer the ground provides both of these.  Additional observations would allow us to identify variable stars, which could affect the LFs of the luminous AGB population.  On the analysis side, we chose a fairly crude set of basis functions to model the $K$ LFs, yet the effects of degeneracy between our model components still remain.  A better analysis would quantify more explicitly the true age and metallicity resolution of near-infrared photometry of luminous evolved stars.  Finally, there remains plenty of progress to be made on stellar models of the RGB and AGB; the work by Marigo et al.\ (2003) and Cioni et al.\ (2005) are a sign of good things to come.

\acknowledgments
We thank Sidney Wolff for her careful reading of an early draft and insightful comments that affected all aspects of this paper.  We thank the anonymous referee for thoughtful comments that clarified important issues.  KO thanks Dara Norman and Tyra Olsen for their support while this work was being done.

\clearpage

\begin{deluxetable}{lcccc}
\tablecaption{Photometric Zero Points} 
\tablewidth{0pt}
\tabletypesize{\small}
\tablehead{ 
\colhead{} & \multicolumn{2}{c}{Star with 1 ADU/s} & \multicolumn{2}{c}{Guide star - Standard zero point} \\
\colhead{Field} & 
\colhead{$H$} &
\colhead{$K$} &
\colhead{$\Delta H-K$} &
\colhead{$\Delta K$}
}
\startdata
Bulge 1& 23.92(0.02) & 23.27(0.03)&  0.04(0.11)  &  -0.04(0.06)\\
Bulge 2& 23.92(0.02) & 23.27(0.03)& -0.07(0.11)  &   0.14(0.06)\\
Disk 2 & 23.92(0.02) & 23.27(0.03)& -0.03(0.11)  &  -0.27(0.06)\\
\enddata

\end{deluxetable}

\begin{deluxetable}{lc}
\tablecaption{50\% Completeness Limits} 
\tablewidth{0pt}
\tabletypesize{\small}
\tablehead{ 
\colhead{Field} & 
\colhead{$K$} 
}
\startdata
Bulge 1& 19.5 \\
Bulge 2& 20.5 \\
Disk 2 & 20.5 \\
F1 & 19.0 \\
F2 & 21.0 \\
F3 & 20.5 \\
F4 & 21.0 \\
F5 & 20.0 \\
F170 & 20.5 \\
F174 & 19.5 \\
F177 & 19.5 \\
F280 & 21.0 \\
\enddata 
\end{deluxetable}

\begin{deluxetable}{lccccccccccc}
\tablecolumns{12}
\tablenum{3}
\tablecaption{Summary of Luminosity Function Fits} 
\tablewidth{0pt}
\tabletypesize{\scriptsize}
\tablehead{ 
\colhead{} & \colhead{} & \colhead{} & \multicolumn{6}{c}{SFR ($M_\odot$/yr$^{-1}$)} \\
\cline{4-9}\\
\colhead{ID} & \colhead{$A_K$} & \colhead{Age} & \colhead{$Z=0.0001$} & \colhead{$0.0004$} & \colhead{$0.001$} & \colhead{$0.008$} & \colhead{$0.019$} & \colhead{$0.03$} & \colhead{$\chi^2_\lambda$} & \colhead{$\sigma_{\chi^2_\lambda}$} & \colhead{$P_\lambda(\%)$} \\
}
\startdata
   Disk2-1 &    0.1 &   1 &                      0.00 &                      0.00 &                      0.00 &                      0.00 &                      0.00 &                      0.00 &    30.65 &     0.59 &  25.69 \\
 & & &                    (0.00) &                    (0.00) &                    (0.00) &                    (0.00) &                    (0.00) &                    (0.00) &  & \\
 & &   3 &                      0.49 &                      0.00 &                      0.00 &                      1.64 &                      0.00 &                      0.00 &  & \\
 & & &                    (0.31) &                    (0.01) &                    (0.00) &                    (0.62) &                    (0.00) &                    (0.00) &  & \\
 & &   5 &                      0.00 &                      0.00 &                      0.00 &                      0.00 &                      0.00 &                      0.95 &  & \\
 & & &                    (0.00) &                    (0.00) &                    (0.00) &                    (0.01) &                    (0.01) &                    (1.16) &  & \\
 & &  10 &                      0.00 &                      0.90 &                      2.32 &                      0.00 &                      0.00 &                      3.58 &  & \\
 & & &                    (0.00) &                    (1.19) &                    (1.29) &                    (0.10) &                    (0.25) &                    (1.47) &  & \\
\enddata
\end{deluxetable}

\begin{deluxetable}{lcccccccccc}
\tablecolumns{11}
\tablecaption{Spearman $r_s$ Matrix} 
\tablewidth{0pt}
\tabletypesize{\small}
\tablehead{ 
\colhead{} & \multicolumn{10}{c}{Component} \\
\colhead{} & \multicolumn{4}{c}{Age (Gyr)} & \multicolumn{6}{c}{$Z$} \\
\cline{2-11}\\
\colhead{Component} & \colhead{10} & \colhead{5} & \colhead{3} & \colhead{1} & \colhead{$0.0004$} & \colhead{$0.0004$} & \colhead{$0.001$} & \colhead{$0.008$} & \colhead{$0.02$} & \colhead{$0.03$} \\
}
\startdata
   10    &           1.00 &  -0.92 &  -0.02 &   0.06 &  -0.08 &   0.11  &  0.13  &  0.47 &   0.28 &  -0.57\\
    5    &                &   1.00 &  -0.26 &  -0.12 &  -0.26 &  -0.18  & -0.04  & -0.36 &  -0.08 &   0.42\\
    3    &                &        &   1.00 &  -0.29 &   0.57 &   0.42  &  0.06  &  0.23 &  -0.36 &  -0.04\\
    1    &                &        &        &   1.00 &   0.47 &  -0.46  & -0.37  & -0.58 &   0.35 &  -0.18\\
 0.0001  &                &        &        &        &   1.00 &  -0.16  & -0.38  & -0.28 &  -0.11 &   0.02\\
 0.0004  &                &        &        &        &        &   1.00  &  0.13  &  0.36 &  -0.71 &   0.20\\
  0.001  &                &        &        &        &        &         &  1.00  &  0.26 &  -0.05 &  -0.22\\
  0.008  &                &        &        &        &        &         &        &  1.00 &  -0.09 &  -0.39\\
    0.02 &                &        &        &        &        &         &        &       &   1.00 &  -0.66\\
    0.03 &                &        &        &        &        &         &        &       &        &   1.00\\
\enddata 
\end{deluxetable}

\begin{deluxetable}{lcccccccccc}
\tablecolumns{11}
\tablecaption{Significance Matrix} 
\tablewidth{0pt}
\tabletypesize{\small}
\tablehead{ 
\colhead{} & \multicolumn{10}{c}{Component} \\
\colhead{} & \multicolumn{4}{c}{Age (Gyr)} & \multicolumn{6}{c}{$Z$} \\
\cline{2-11}\\
\colhead{Component} & \colhead{10} & \colhead{5} & \colhead{3} & \colhead{1} & \colhead{$0.0004$} & \colhead{$0.0004$} & \colhead{$0.001$} & \colhead{$0.008$} & \colhead{$0.02$} & \colhead{$0.03$} \\
}
\startdata
   10    &           0.00 &   0.00 &   0.95 &   0.85 &   0.81 &   0.74 &   0.68 &   0.12 &   0.38 &   0.05\\
    5    &                &   0.00 &   0.42 &   0.71 &   0.41 &   0.57 &   0.91 &   0.25 &   0.81 &   0.17\\
    3    &                &        &   0.00 &   0.35 &   0.06 &   0.17 &   0.86 &   0.48 &   0.26 &   0.90\\
    1    &                &        &        &   0.00 &   0.12 &   0.14 &   0.23 &   0.05 &   0.27 &   0.57\\
 0.0001  &                &        &        &        &   0.00 &   0.62 &   0.22 &   0.38 &   0.74 &   0.95\\
 0.0004  &                &        &        &        &        &   0.00 &   0.69 &   0.25 &   0.01 &   0.53\\
  0.001  &                &        &        &        &        &        &   0.00 &   0.41 &   0.88 &   0.49\\
  0.008  &                &        &        &        &        &        &        &   0.00 &   0.78 &   0.20\\
    0.02 &                &        &        &        &        &        &        &        &   0.00 &   0.02\\
    0.03 &                &        &        &        &        &        &        &        &        &   0.00\\
\enddata 
\end{deluxetable}

\clearpage

\begin{figure}
\caption{The locations of the NIRI/Altair fields studied in Paper I and the HST/NICMOS fields presented by S03 are shown with respect to the $R$-band Local Group Survey image of M31 (Massey et al. 2002).\label{image}}
\end{figure}
\begin{figure}
\caption{Aperture corrections as a function of radius.  The difference between the magnitudes measured within a 15-pixel radius aperture and the PSF magnitudes  of the stars used to define the PSF in our Bulge 2 $K$ image is plotted vs. radius from the central guide star.  Anisoplanaticity produces a strong radial dependence of the aperture correction.\label{apcor}}
\end{figure}
\begin{figure}
\caption{The new $H-K,K$ color-magnitude diagrams of our three NIRI/Altair fields are shown here. We attribute the small differences with the CMDs of Paper 1 to the inclusion of aperture corrections as a function of radius in the present work.\label{cmds}}
\end{figure}
\begin{figure}
\caption{$H-K$ vs. radius from the central guide star in our Bulge 1 NIRI/Altair field.  The dotted line is a fit to the points, and demonstrates that the mean $H-K$ color is independent of radius in our photometry.  The dashed line is a fit to the mean color vs. radius of our Paper I photometry, which contains residual effects from uncorrected anisoplanaticity. \label{hkr}}
\end{figure}
\begin{figure}
\caption{Simulated M31 bulge and disk stellar populations as seen by NIRI/Altair.  Each simulation was produced by inserting an appropriate number of stars drawn from a 12-Gyr solar metallicity G02 isochrone into artificial images, convolving the images with our DAOPHOT/ALLSTAR PSFs, adding noise, and running the images through our photometric reduction pipeline.  The photometric depths and scatter at faint magnitudes closely matches that which we observe in our fields; see text.  At bright magnitudes, the scatter in color is much smaller than observed, demonstrating that the M31 fields contain a mix of stellar populations. \label{simcmds}}
\end{figure}
\begin{figure}
\caption{Photometric completeness, error, and bias in our Bulge 2 NIRI/Altair field as calculated by comparing the input and recovered photometry of 20000 artificial stars inserted into our observed images.  {\it Top left:} Completeness versus $K$ magnitude.  {\it Top right:} Photometric error in $K$ (solid circles) and $H-K$ (open circles) vs. $K$; overplotted are our analytical predictions using the equations from Olsen, Blum, \& Rigaut (2003), with dotted line designating the $K$ and dashed line the $H-K$ errors. The fact that the $H-K$ errors are smaller than the $K$ errors indicates that crowding is the dominant source of error.  {\it Bottom left:} The input minus recovered $K$ magnitudes of our artificial stars.  Below $\sim$50\% completeness, the photometric bias becomes large.  {\it Bottom right:} As in bottom left, but for $H-K$. None of these quantities vary appreciably with radius from the AO guide star within the range of radii to which our photometry is restricted.  \label{errfig}}
\end{figure}

\begin{figure}
\caption{{\it Left:} Comparison of NIRI/Altair (red) and HST/NICMOS (black) $(H-K)_0, M_K$ color-magnitude diagrams combined over all fields.  We have assumed $A_K=0.1$ and $(m-M)_0$=24.45 in constructing the CMDs.  {\it Right:} $(H-K)_0$ color distributions for all stars within 0.25 magnitudes of $M_K=-6$ for the NIRI/Altair (red) and HST/NICMOS (black) combined CMDs, normalized by the total number of stars in the distributions.  The NIRI/Altair distribution lies 0.037 magnitudes to the blue of the HST/NICMOS distribution, and has scatter of 0.081 magnitudes compared to 0.064 for HST/NICMOS.\label{nicvsniri}}
\end{figure}

\begin{figure}
\caption{{\it Left:} Comparison of 2MASS PSC $(H-K)_0, M_K$ CMD of a region in the LMC Bar (red) with the Stephens et al. (2003) HST/NICMOS CMD (black) of all M31 fields combined.  We have assumed $A_K=0.1$ and $(m-M)_0$=24.45 in constructing the M31 CMD, and $A_K$=0.025 and $(m-M)_0$=18.5 for the LMC CMD.  {\it Right:} $(H-K)_0$ color distributions for all stars within 0.25 magnitudes of $M_K=-6$ for the 2MASS LMC Bar (red) and HST/NICMOS M31 (black) combined CMDs, normalized by the total number of stars in the distributions.  The LMC Bar distribution lies 0.1 magnitudes to the blue of the HST/NICMOS distribution, as expected from the LMC's lower metallicity and younger age distribution.\label{nicvslmchk}}
\end{figure}

\begin{figure}
\caption{As in Figure \ref{nicvslmchk}, but for $(J-K)_0$ instead of $(H-K)_0$.  The LMC Bar color distribution appears $\sim$0.05 magnitudes {\em redder} than the M31 color distribution, contrary to the sense of the $(H-K)_0$ distributions and to our expectations of lower metallicity and younger ages in the LMC, which should have caused the LMC Bar CMD to be $\sim$0.2 magnitudes bluer than the M31 CMD in $(J-K)_0$ at $M_K=-6$. We thus propose that there is an unexplained $\sim$-0.25 magnitude zero point shift in the $J$ photometry of Stephens et al.\ (2003).  \label{nicvslmcjk}}
\end{figure}

\begin{figure}
\caption{An illustration of the sensitivity of the $M_K$ luminosity function to age and metallicity, using the G02 isochrones.  All models have been shifted vertically to match at $M_K=-4$.  {\it Left:} The logarithmic $M_K$ LF is shown for Z= 0.0001 (cyan), 0.0004 (blue), 0.001 (green), 0.008 (yellow), and 0.019 (red) at a fixed age of 5 Gyr.  {\it Right:} $M_K$ LFs are shown for ages of 1 Gyr (blue), 3 Gyr (green), 5 Gyr (yellow), and 10 Gyr (red) at fixed solar metallicity.  \label{modlf}}
\end{figure}

\begin{figure}
\caption{Test \#1, demonstrating our ability to recover a simulated population mix.  The input population box is shown on top left, while our derived solution is shown on top right. In the bottom panels, we show the input (dashed line) and recovered (solid line) age and metallicity distributions. \label{test1}}
\end{figure}
\begin{figure}

\caption{Test \#2, demonstrating our ability to discriminate between 5 and 10-Gyr populations.  The panels are as in Figure \ref{test1}. \label{test2}}
\end{figure}
\begin{figure}
\caption{Test \#3, testing our ability to recover the correct solution in which the input basis set does not match that used in our solution.  The panels are as in Figure \ref{test1}. As seen in bottom left, while we recover the correct age distribution for ages $\le$5 Gyr, all of the star formation occurring at ages $\ge$6 Gyr gets drawn into the 10 Gyr bin. \label{test3}}
\end{figure}
\begin{figure}
\caption{Population boxes derived for our three NIRI/Altair fields and one of the NICMOS fields by fitting our models to the $M_K$ luminosity functions of the data.  In the columns in the middle and on the right, we show the observed (black lines with error bars) and model (red lines) luminosity functions and the residual Hess diagrams; white pixels represent areas where the data are high, while dark pixels are areas where the models predict too many stars. \label{popbox}}
\end{figure}

\begin{figure}
\caption{The stellar populations of M31 as a function of radius.  Solid circles show the results of our fits to the NIRI/Altair and NICMOS $K$ LFs; large open circles identify the three NIRI/Altair fields.  Bulge-to-disk ratios from Kent (1989) are labelled on the upper axes.  {\it Top left:}  Mean age of our population fits, integrated over all metallicities, as a function radius from the M31 center.  Error bars show the rms of the age distributions.  The oldest populations dominate at all radii.  {\it Top right:} The relative contributions of each age component of the models are plotted here.  Correlations between e.g. the 5- and 10-Gyr components hint at partial degeneracy between some of the components.  {\it Bottom left:} Mean metallicities of our population fits, integrated over all ages, as a function radius from the M31 center.  Error bars show the rms of the metallicity distributions.  The bulge is predominantly solar metallicity; the decline in metallicity at radii below 4\arcmin ~is caused entirely by the appearance of an intermediate-age metal-poor component in the fits, which may be spurious (see text).  The disk appears, on average, mbore metal-poor than the bulge. {\it Bottom right:}  The relative contributions of each metallicity component of the models are plotted here.  The components with [M/H]$\ge-0.4$ generally dominate.  \label{radplots}}
\end{figure}

\begin{figure}
\caption{As in Figure \ref{radplots}, but for the fits with $A_K=0.0$ and $A_K=0.2$.  The overall effects of changing $A_K$ are to make the mean metallicities $\sim$0.25 dex lower in the $A_K=0.2$ fits and $\sim$0.25 dex higher in the $A_K=0.0$ fits compared to the case of $A_K=0.1$.  With the exception of the outermost field (F280), the mean ages are $\sim$1 Gyr younger in the case of $A_K=0.2$ while the $A_K=0.0$ mean ages are similar to $A_K=0.1$.  We conclude that our results are insensitive to changes in $A_K$ at the $\sim$0.1 magnitude level.  \label{radplots02}}
\end{figure}

\begin{figure}
\caption{As in Figure \ref{radplots}, but choosing the fits with $A_K$ producing highest statistical likelihood $P_\lambda$. The results are nearly identical to those of Figure \ref{radplots}.  \label{radplots_best}}
\end{figure}
 \clearpage

\begin{figure}
\caption{The metallicity distribution function and population box of the F170 NICMOS field.  Although our metallicity resolution is coarser than Sarajedini \& Jablonka (2005), the distribution functions are in qualitative agreement.  We do not find that all of the populations have ages as old as 10 Gyr, as assumed by Sarajedini \& Jablonka, although these old populations do comprise the biggest component.   \label{sfh170}}
\end{figure}
\begin{figure}
\caption{The integrated population box of M31.  Because of its higher surface brightness, the bulge populations dominate the box.  Although our models consisted of discrete ages from the set \{1, 3, 5, 10\} Gyr and discrete metallicities from the set Z=\{0.0001, 0.0004, 0.001, 0.008, 0.019, 0.03\}, the widths of the cells in age correspond roughly to the age resolution found in our tests.   \label{sfrtot}}
\end{figure}
\begin{figure}
\caption{The star formation rate history and age-metallicity relation, as derived from our LF fits for all of our M31 fields combined, are plotted against those for the simulated disk galaxy of Robertson et al.\ (2004). Filled circles show our results excluding the [M/H]=-2.3 component; open circles includes this metal-poor component.  Our age and metallicity resolution are not yet high enough to favor one numerical simulation over another. \label{simcomp}}
\end{figure}

\clearpage
\plotone{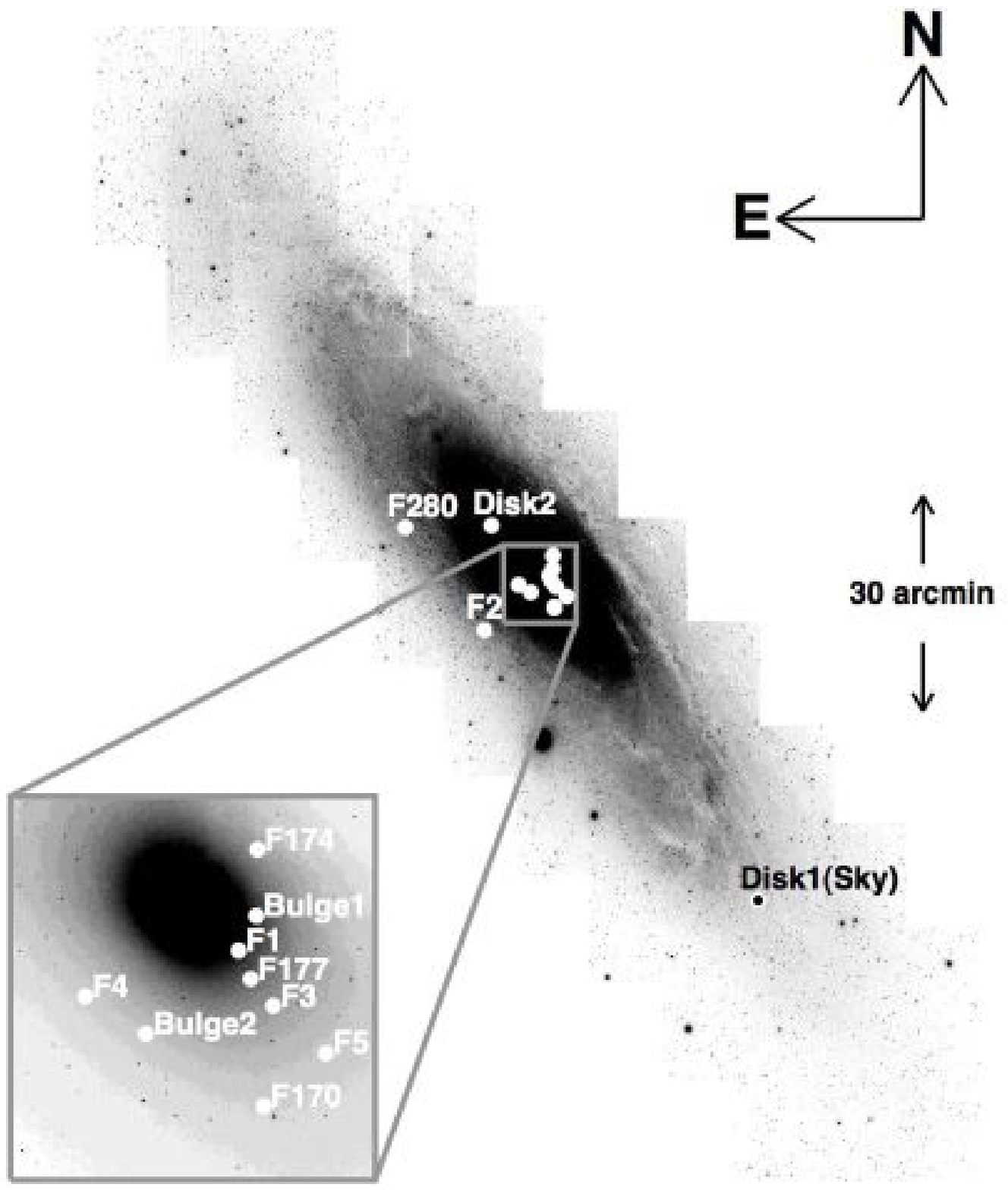}
\newpage
\plotone{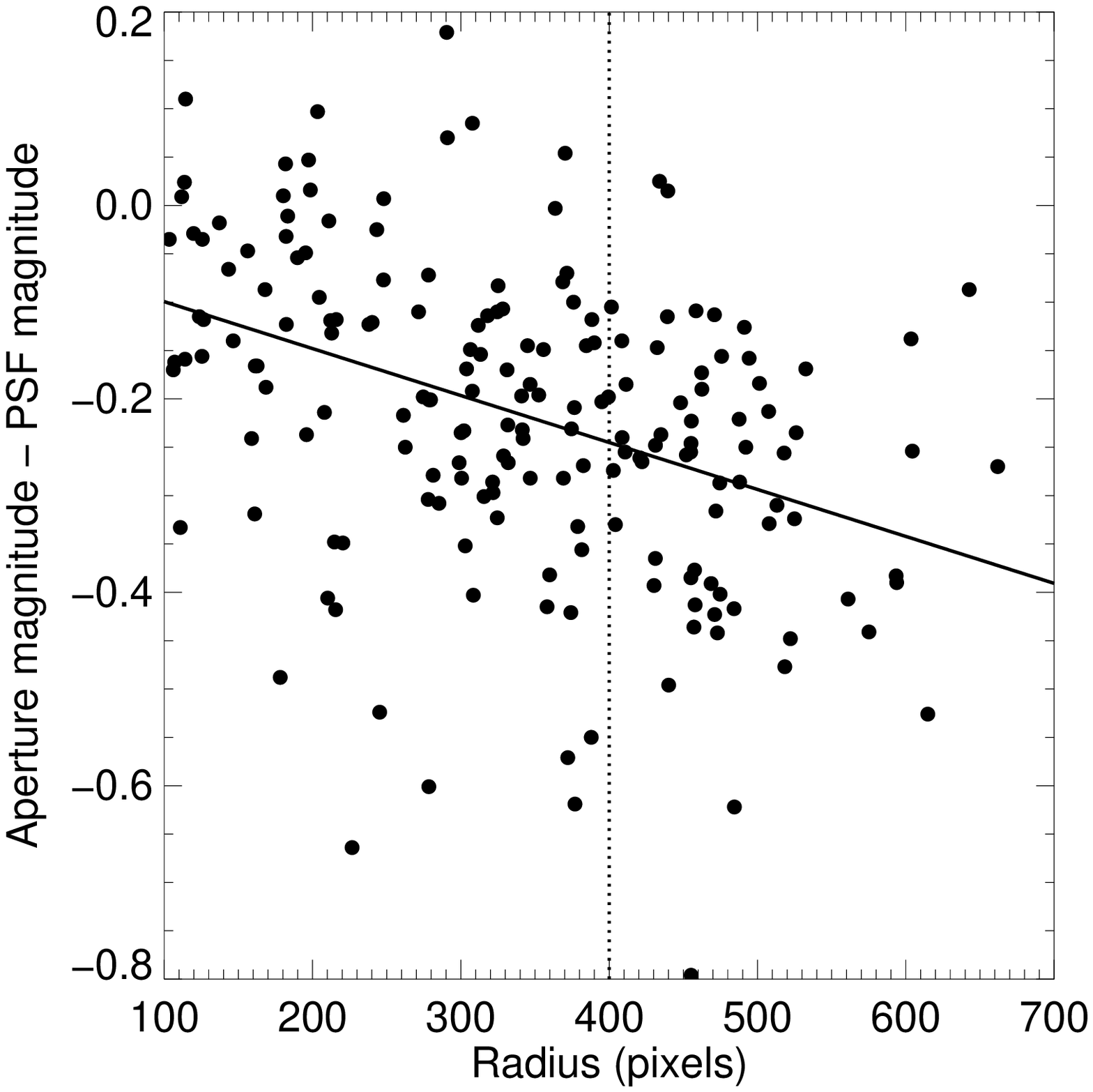}
\newpage
\plotone{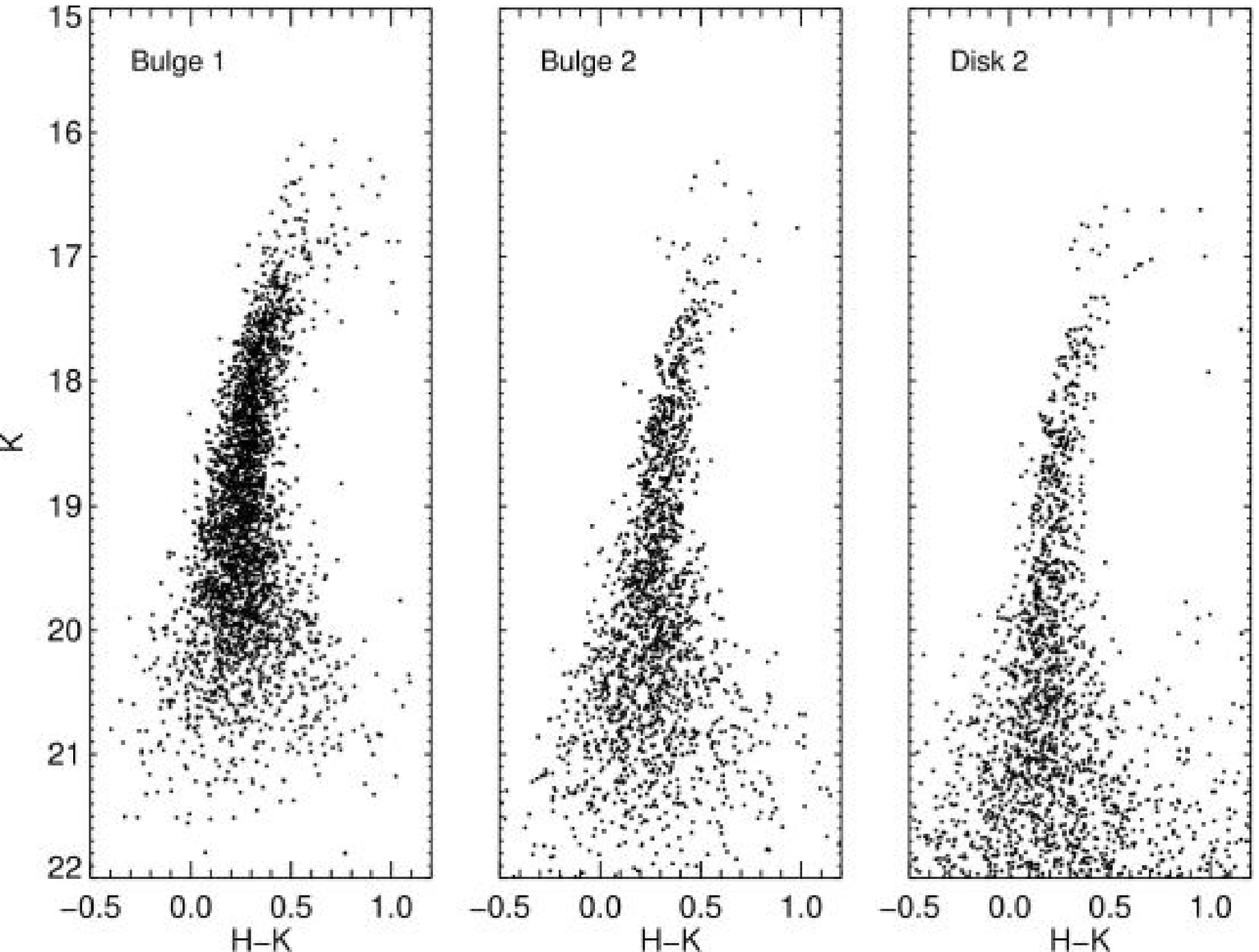}
\newpage
\plotone{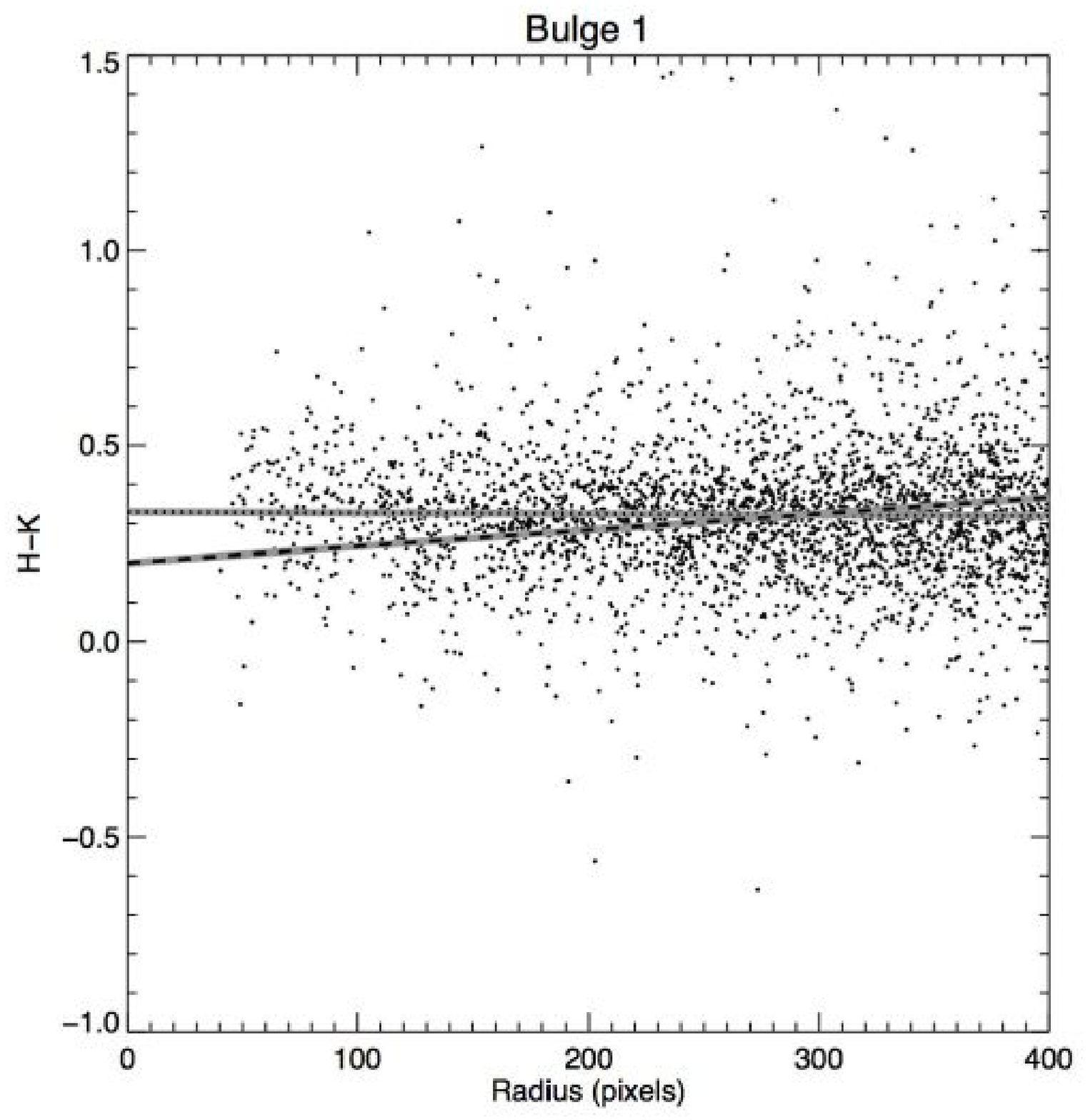}
\newpage
\plotone{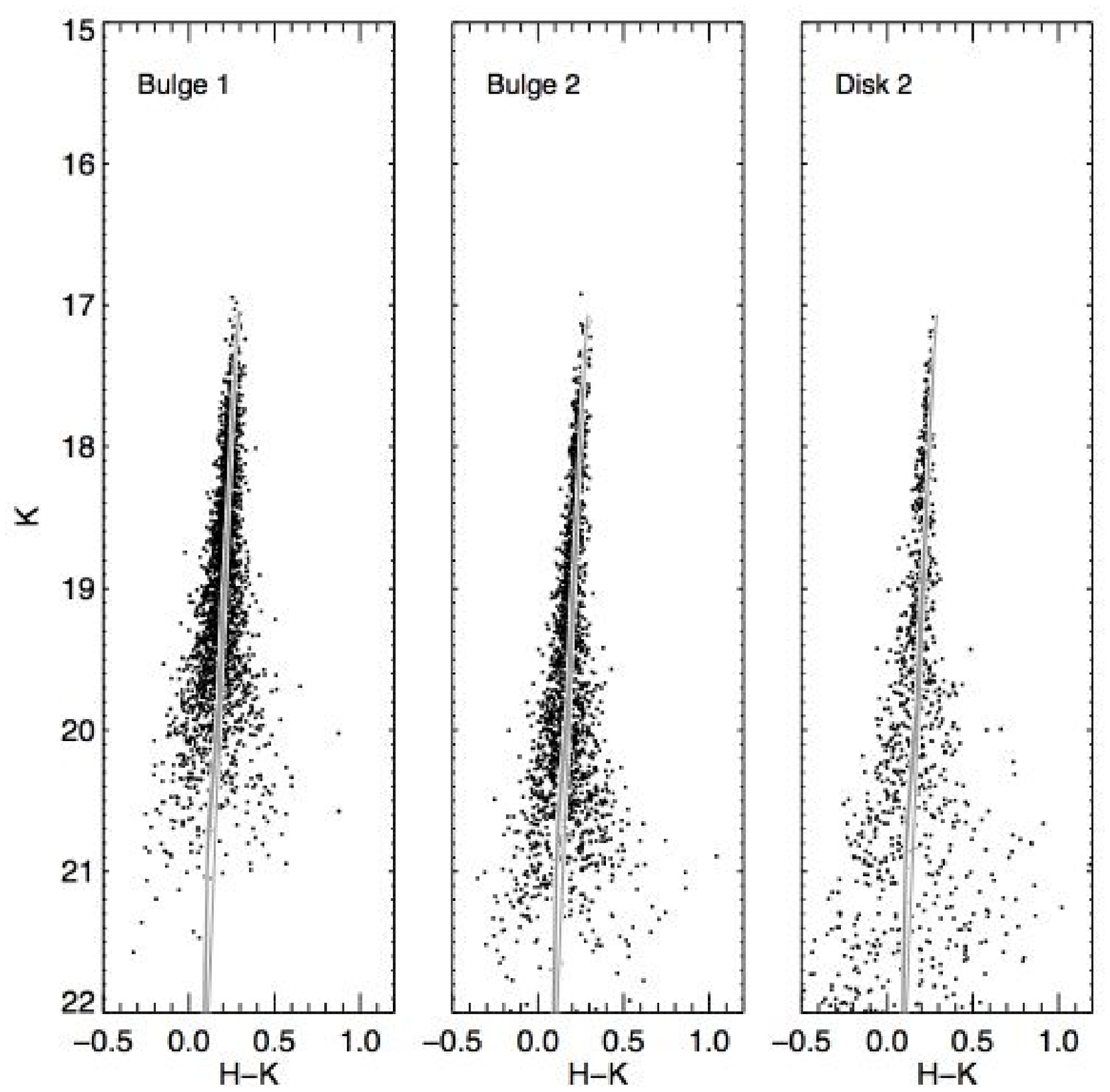}
\newpage
\plotone{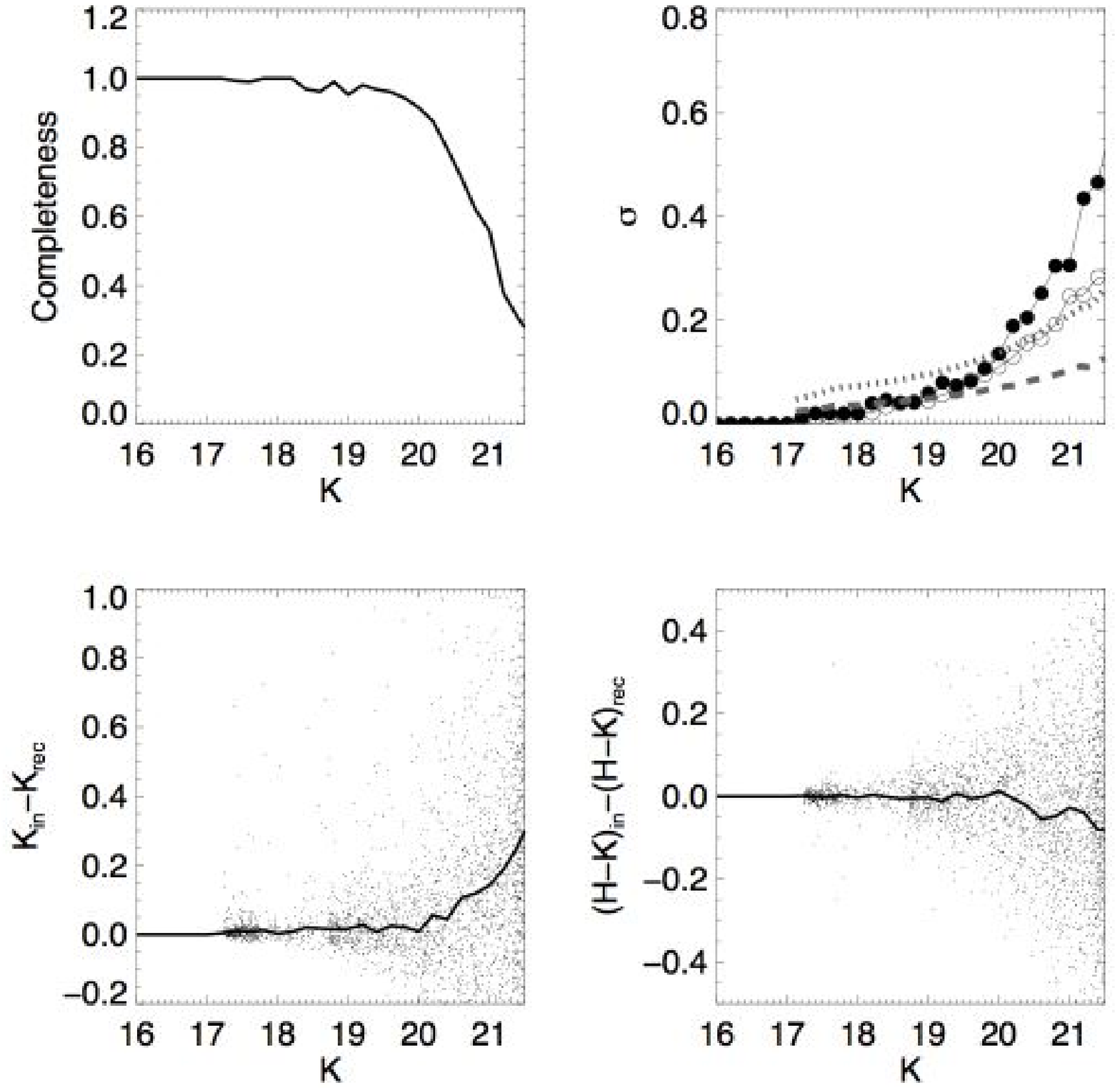}
\newpage
\plotone{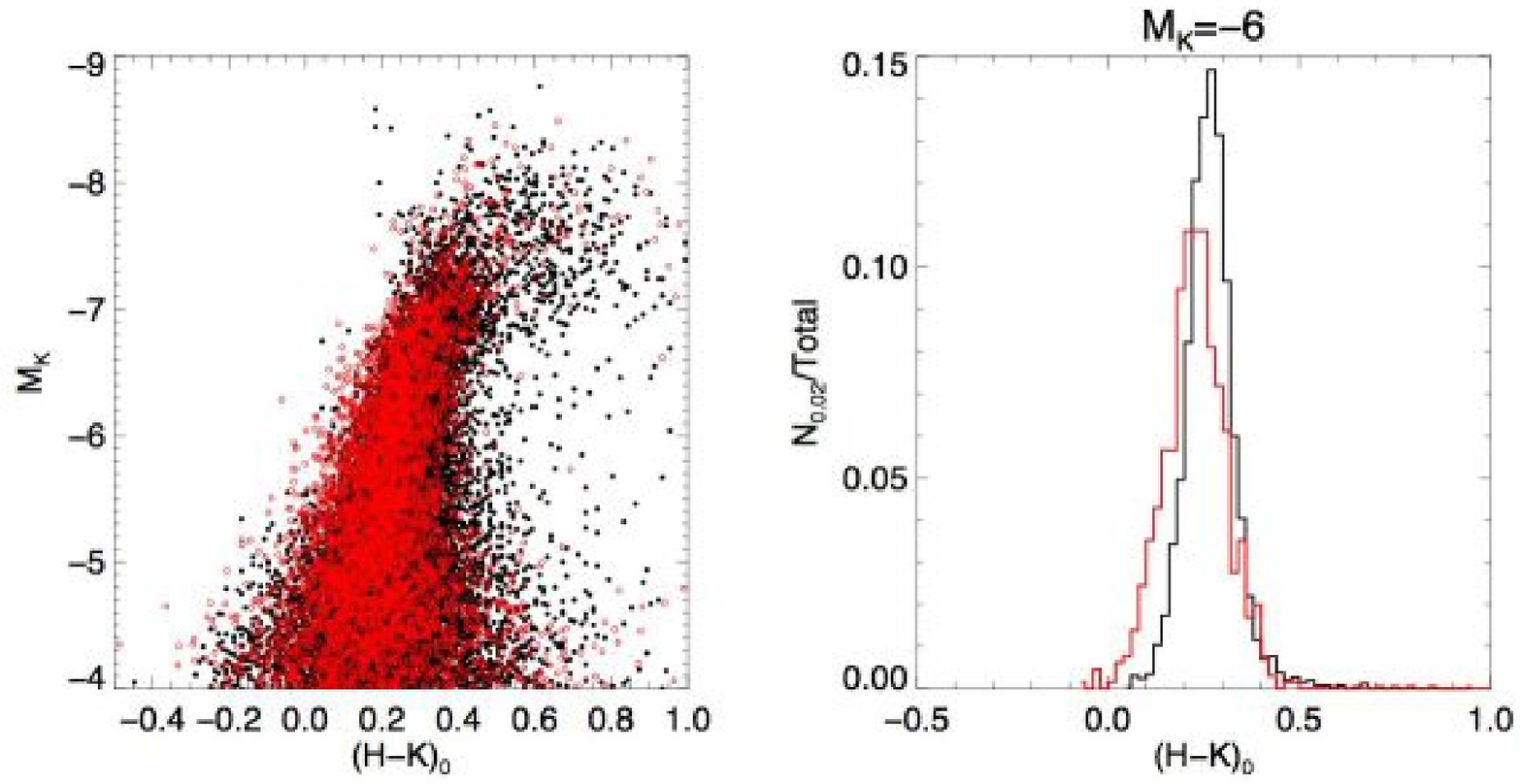}
\newpage
\plotone{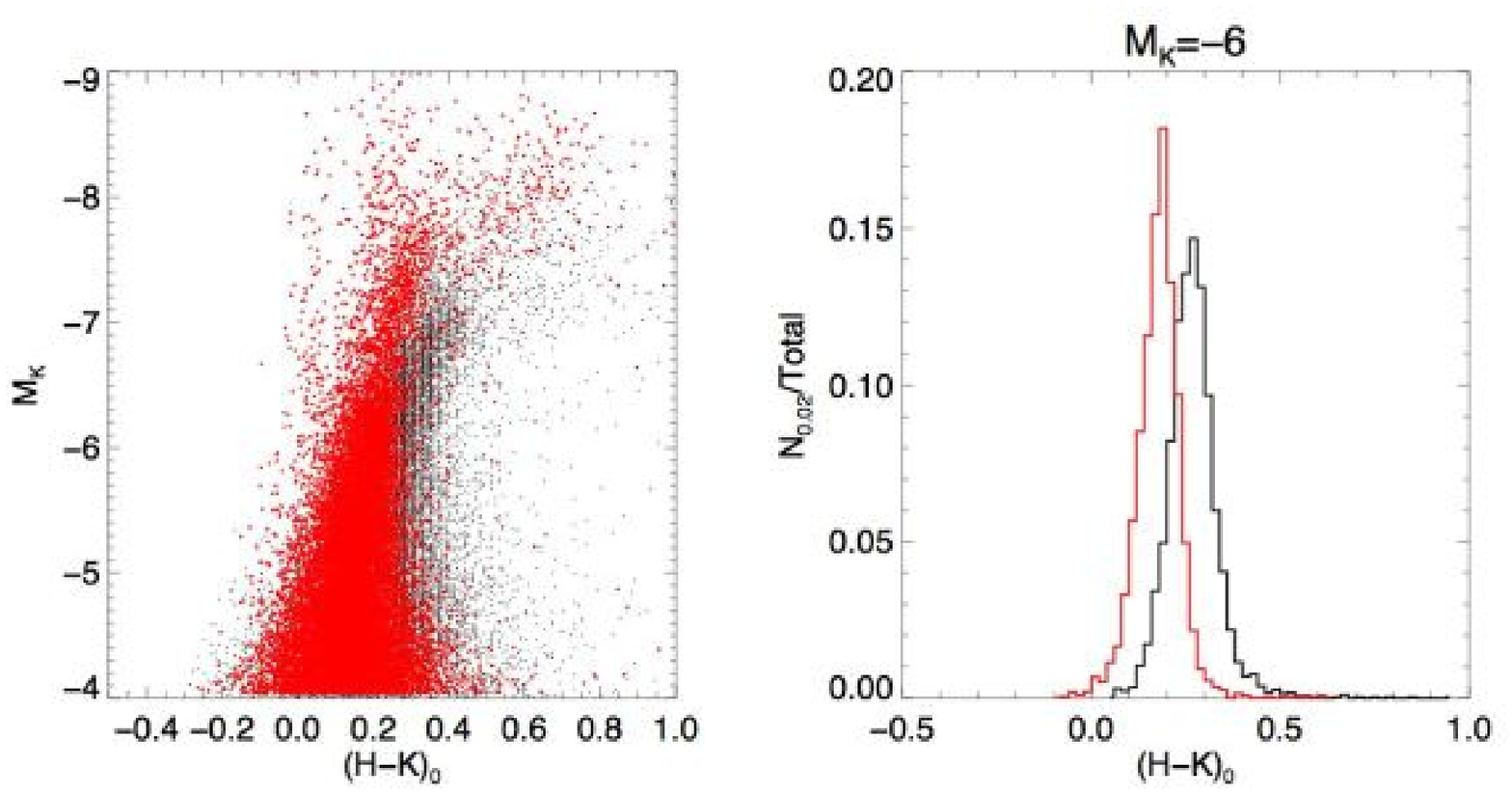}
\newpage
\plotone{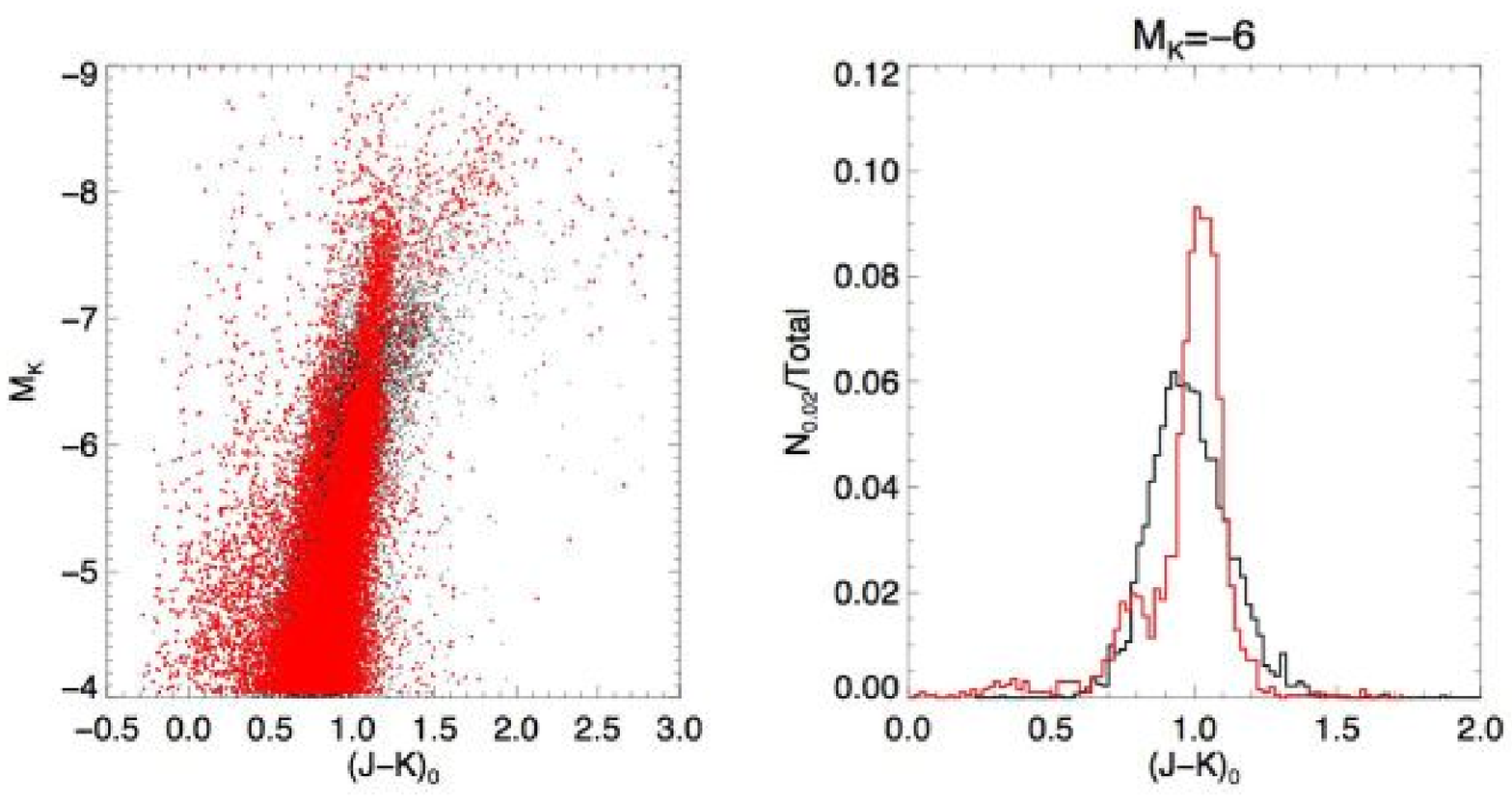}
\newpage
\plotone{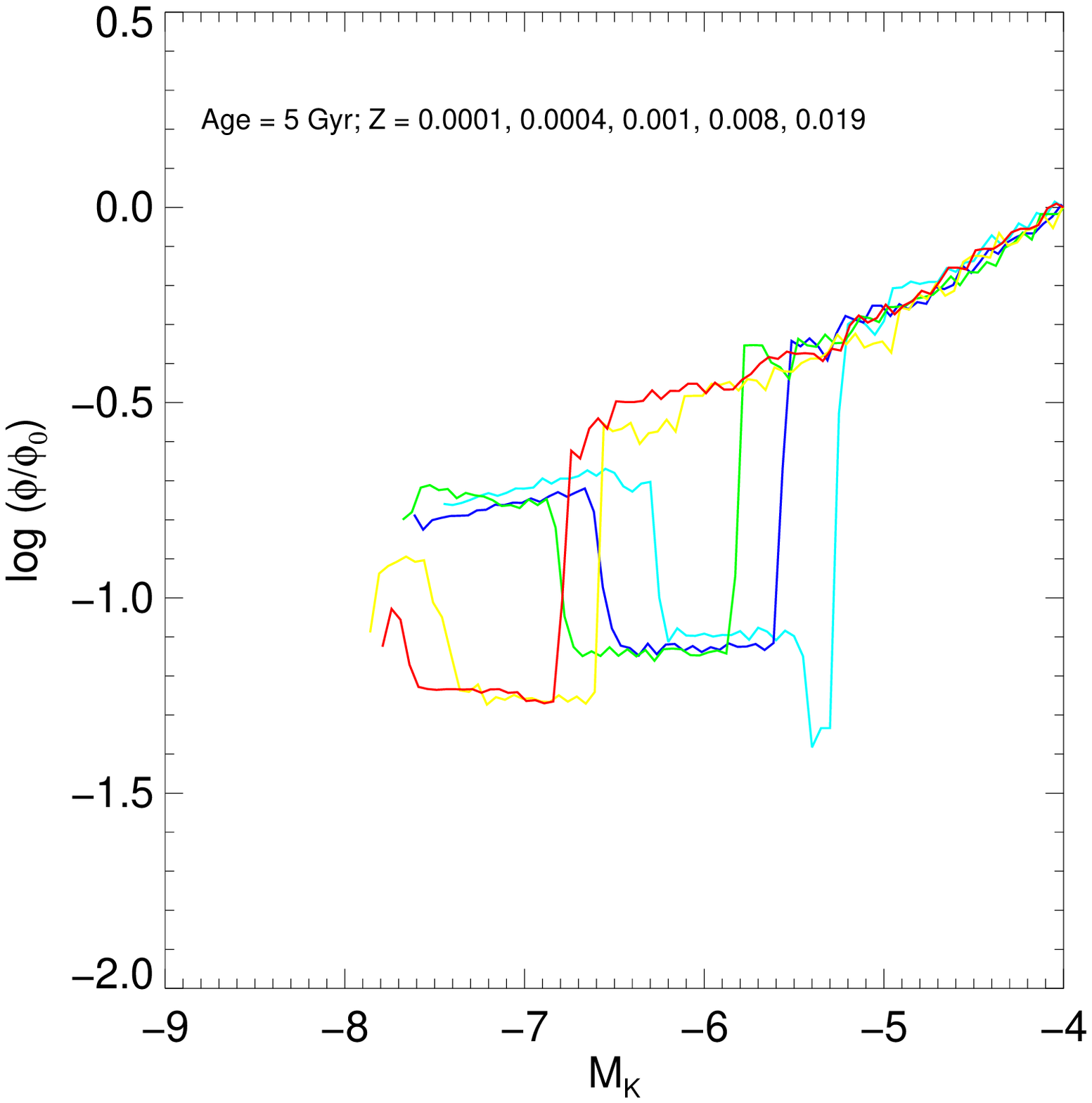}
\newpage
\plotone{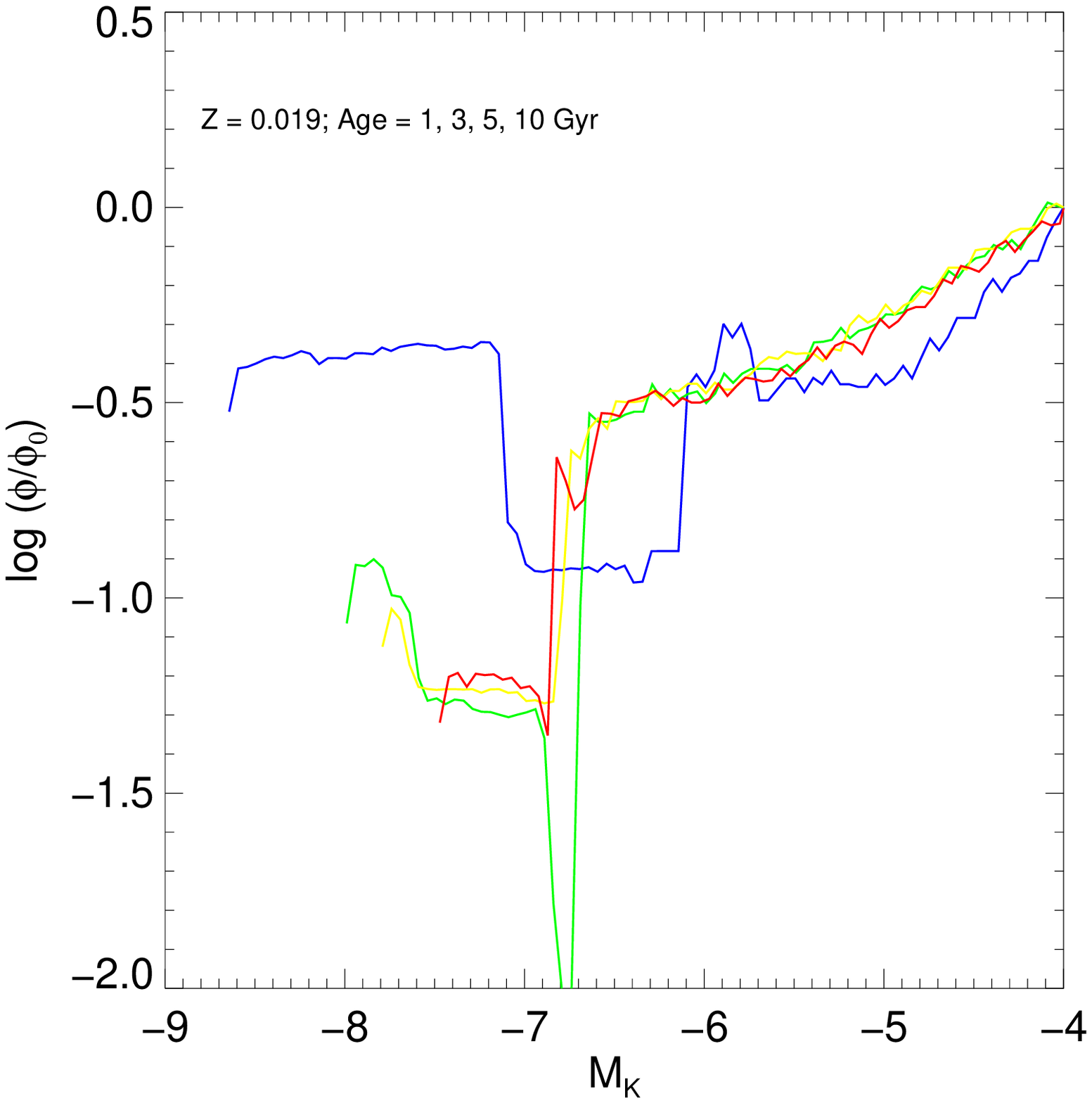}
\newpage
\plotone{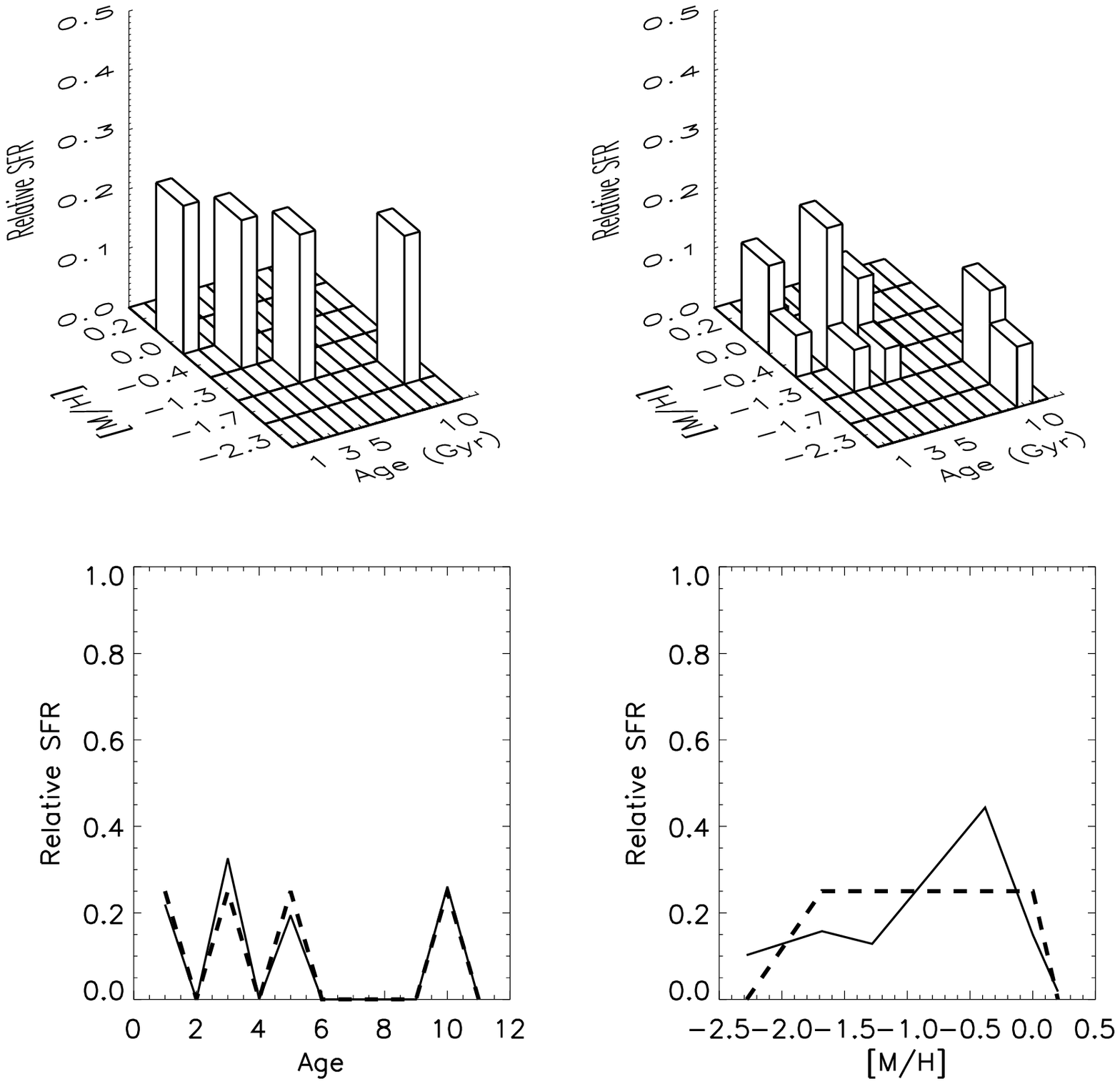}
\newpage
\plotone{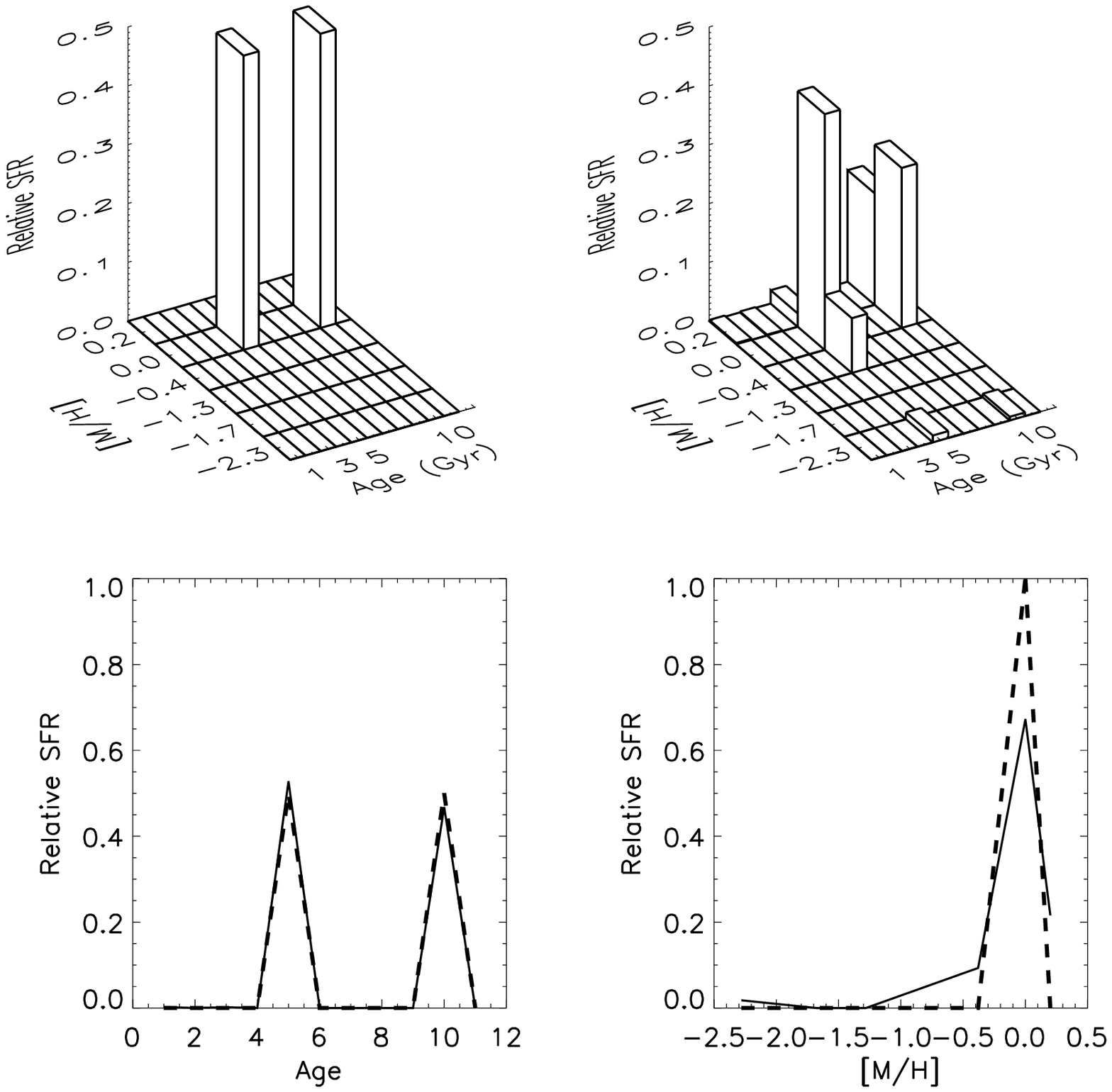}
\newpage
\plotone{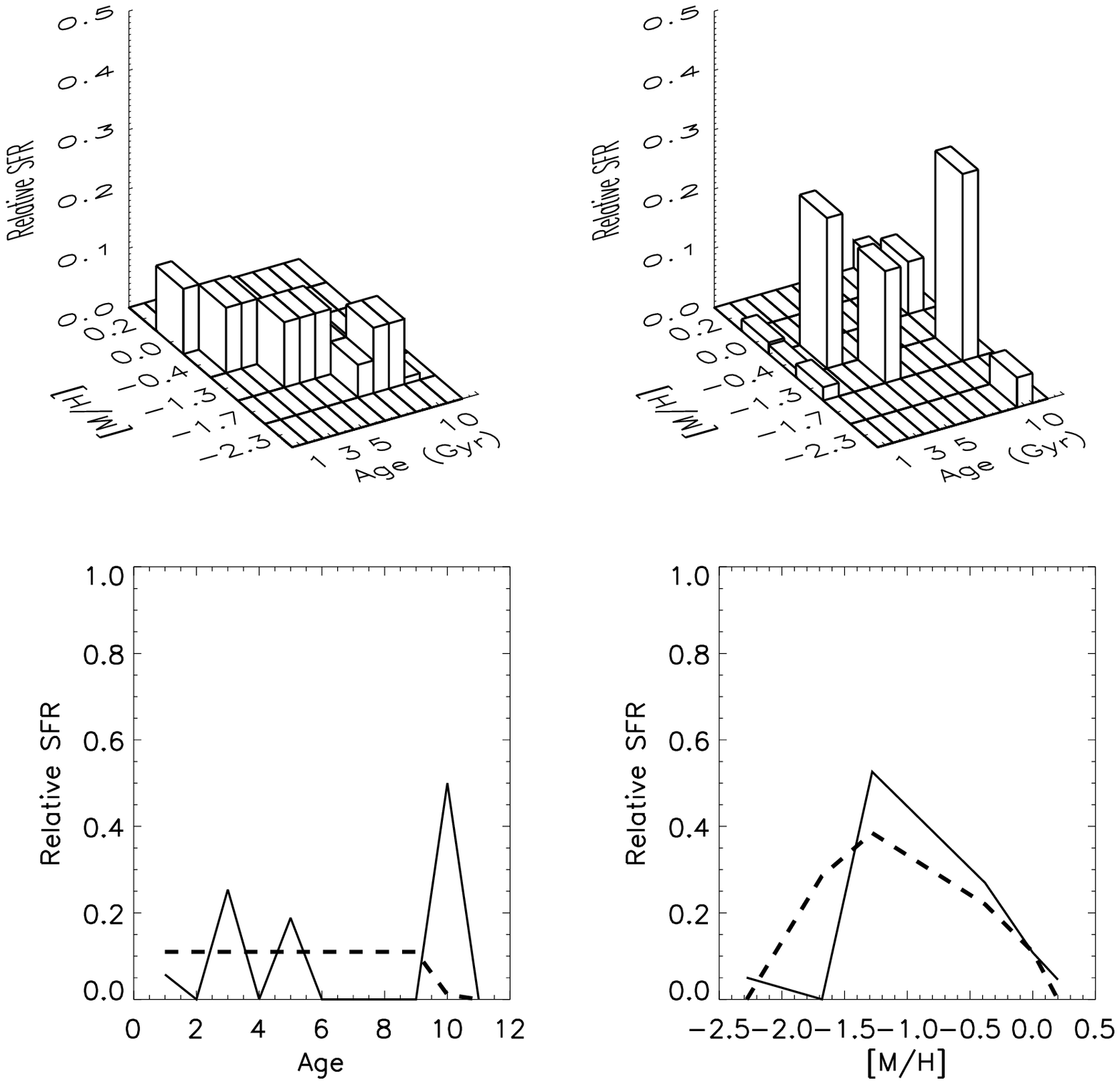}
\newpage
\plotone{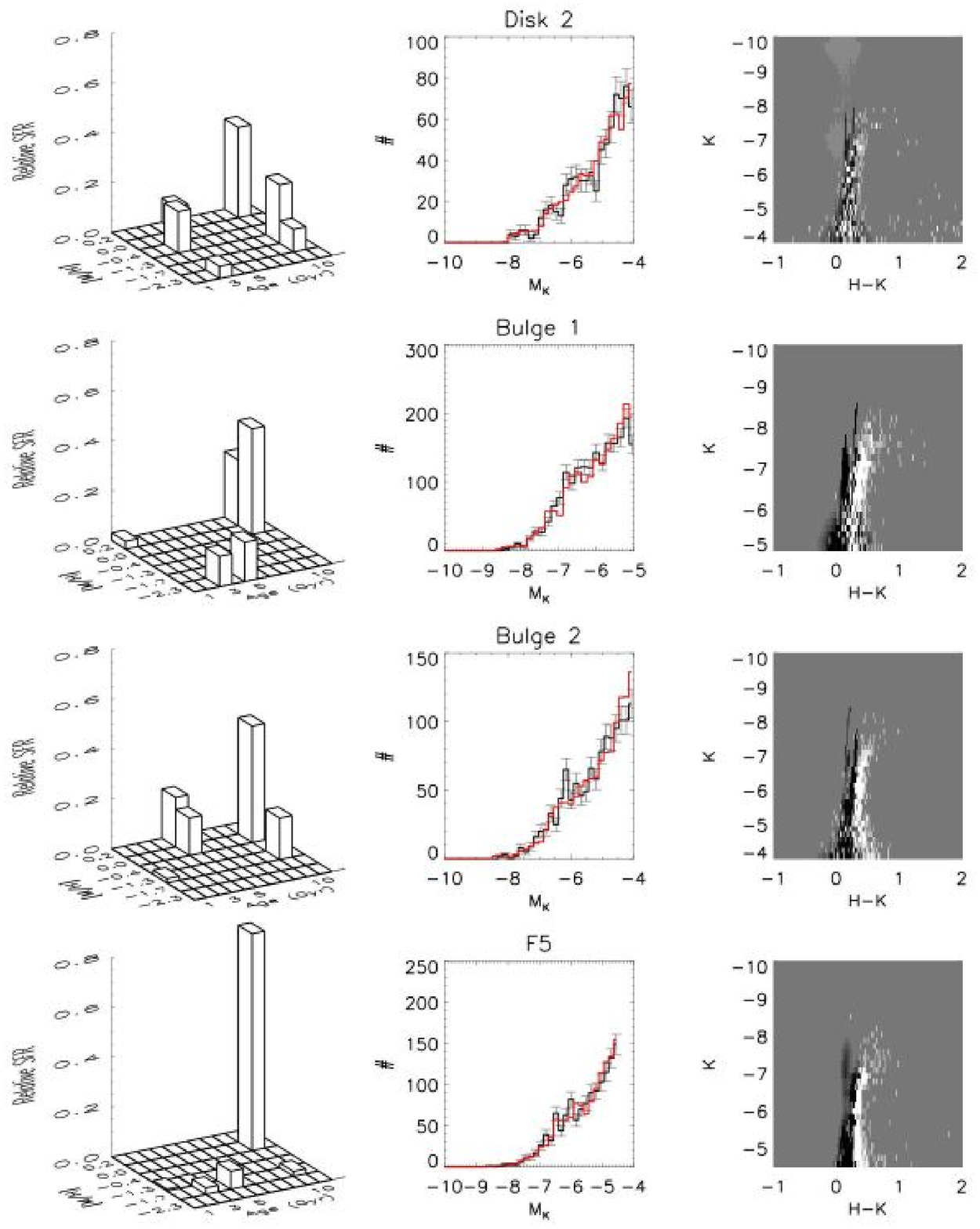}
\newpage
\plotone{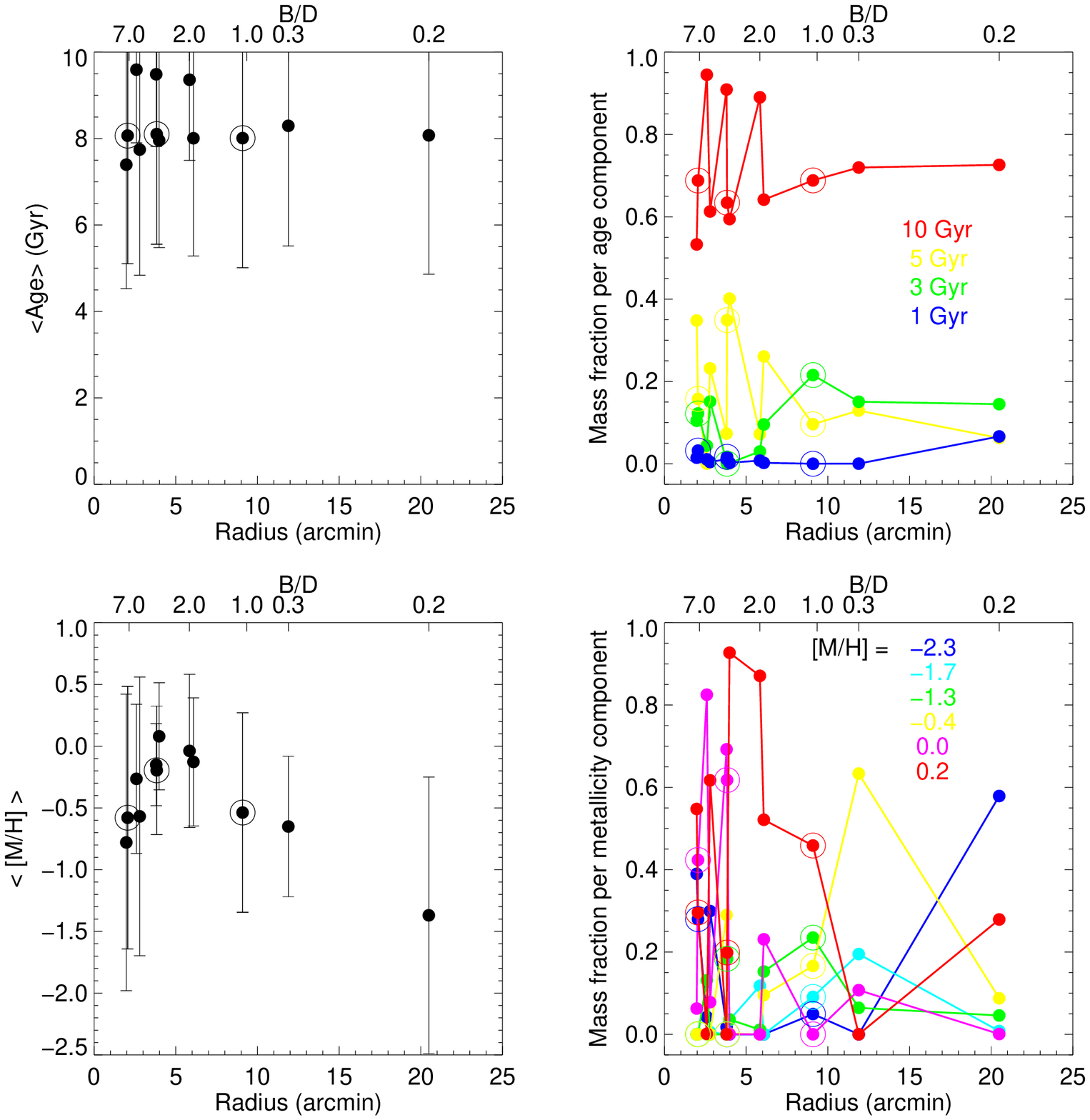}
\newpage
\plotone{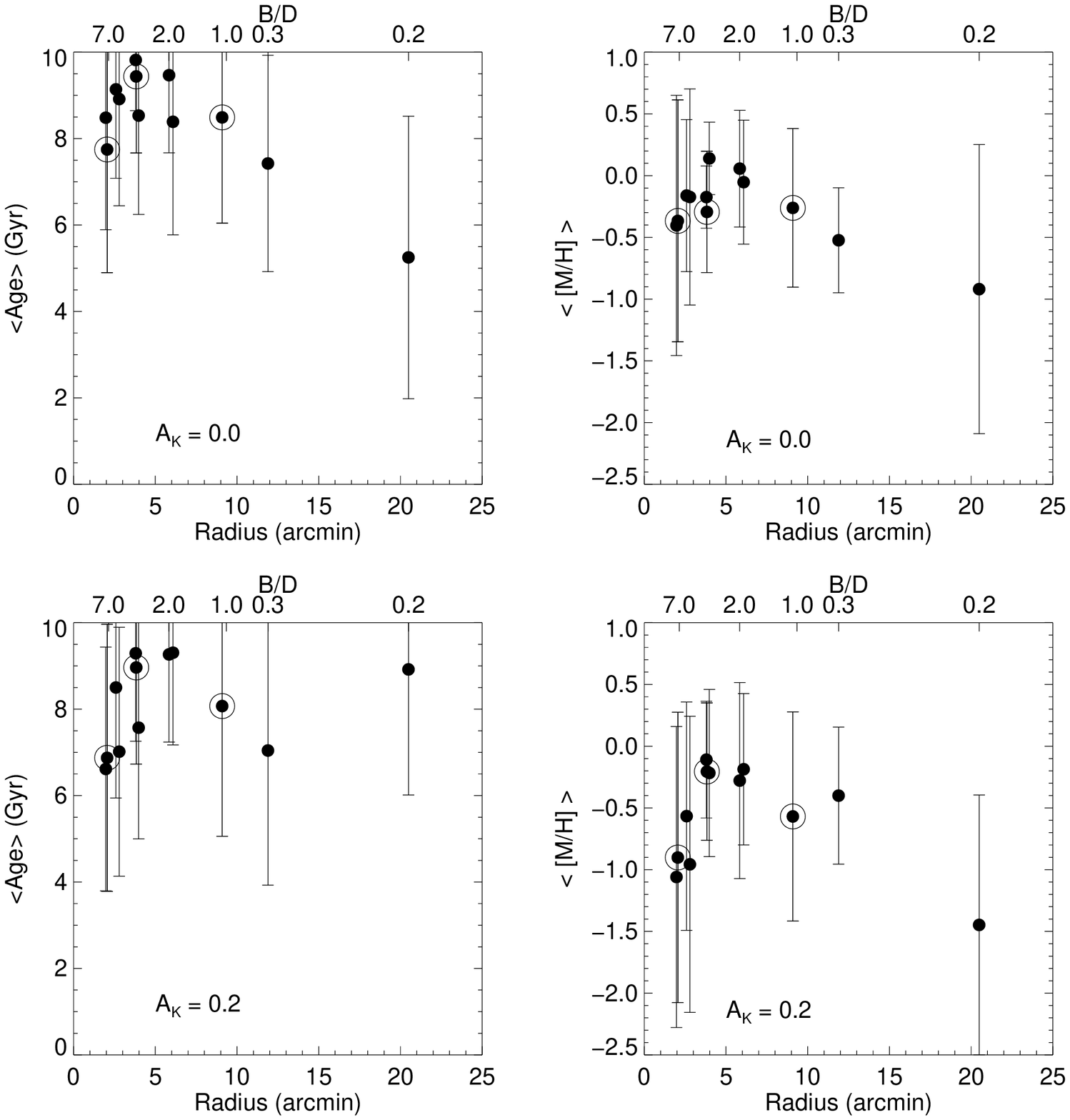}
\newpage
\plotone{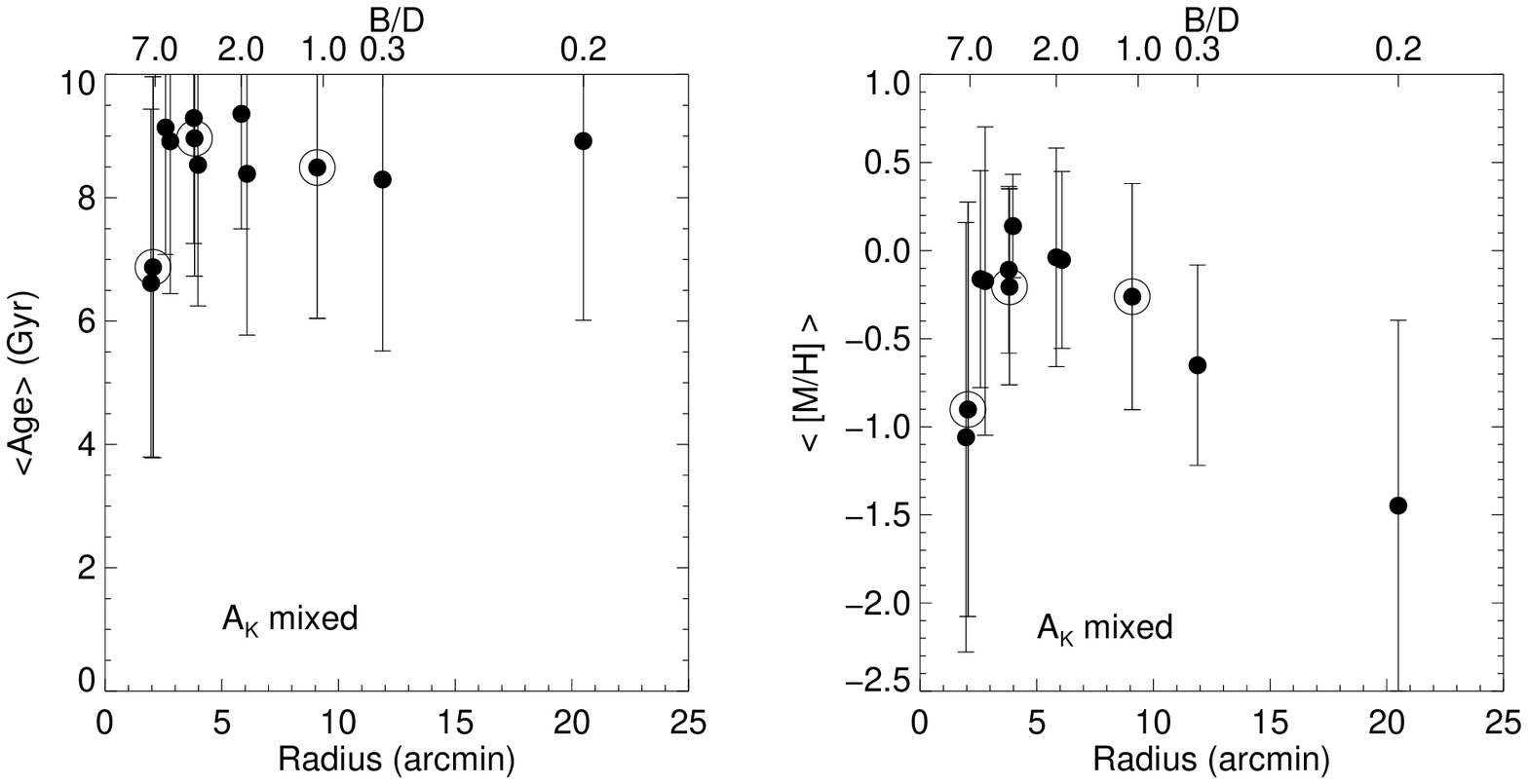}
\newpage
\plotone{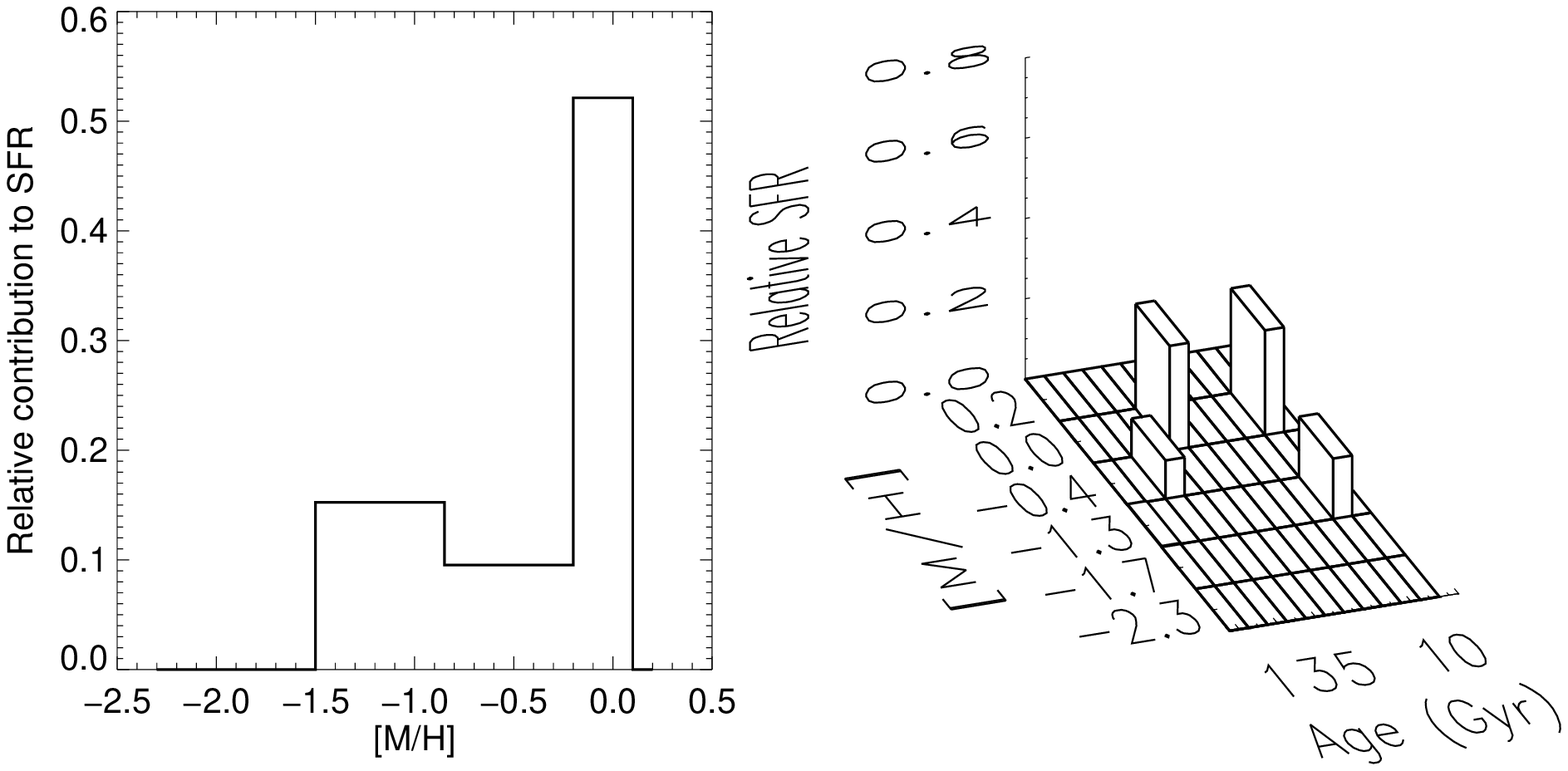}
\newpage
\plotone{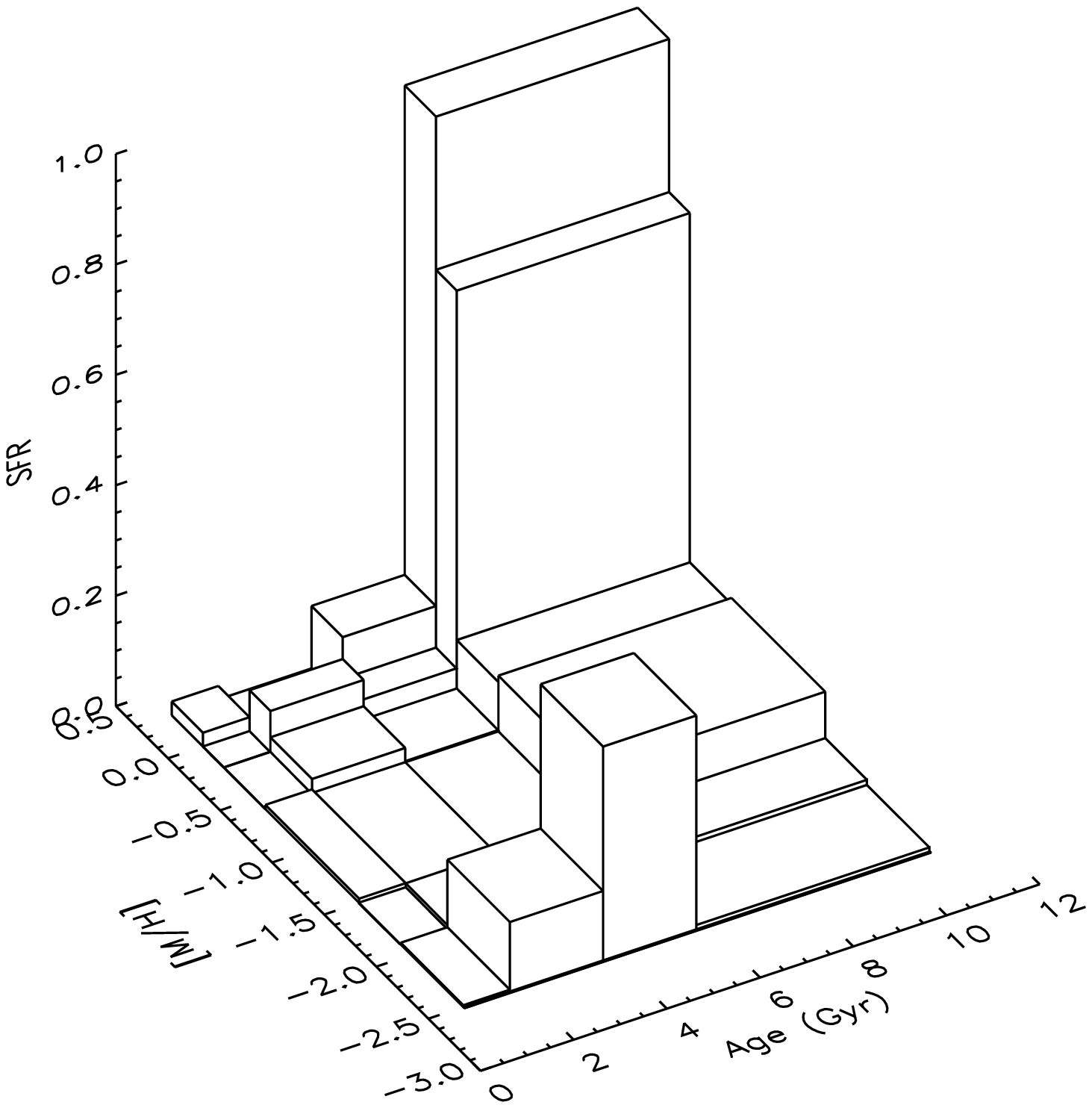}
\newpage
\plotone{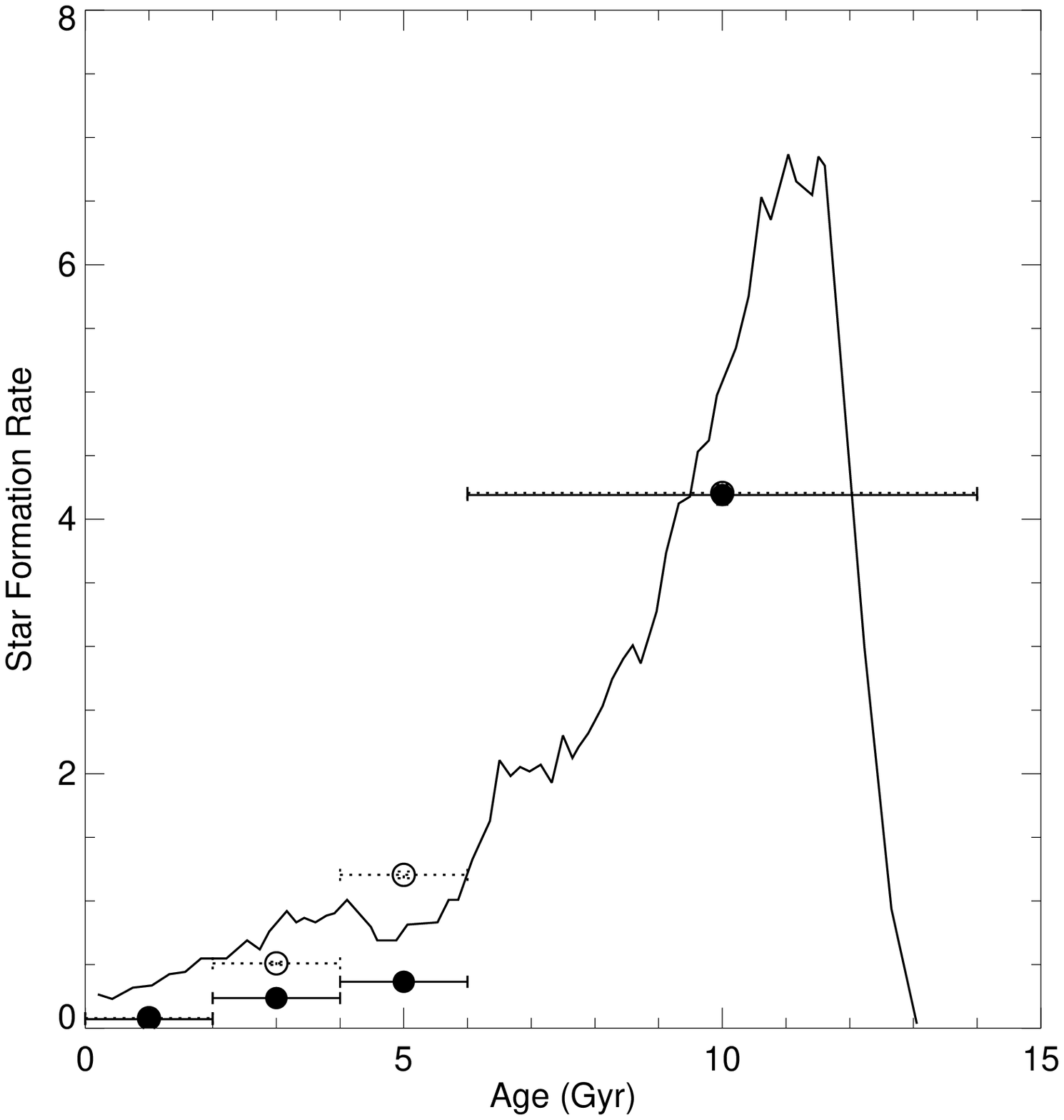}
\newpage
\plotone{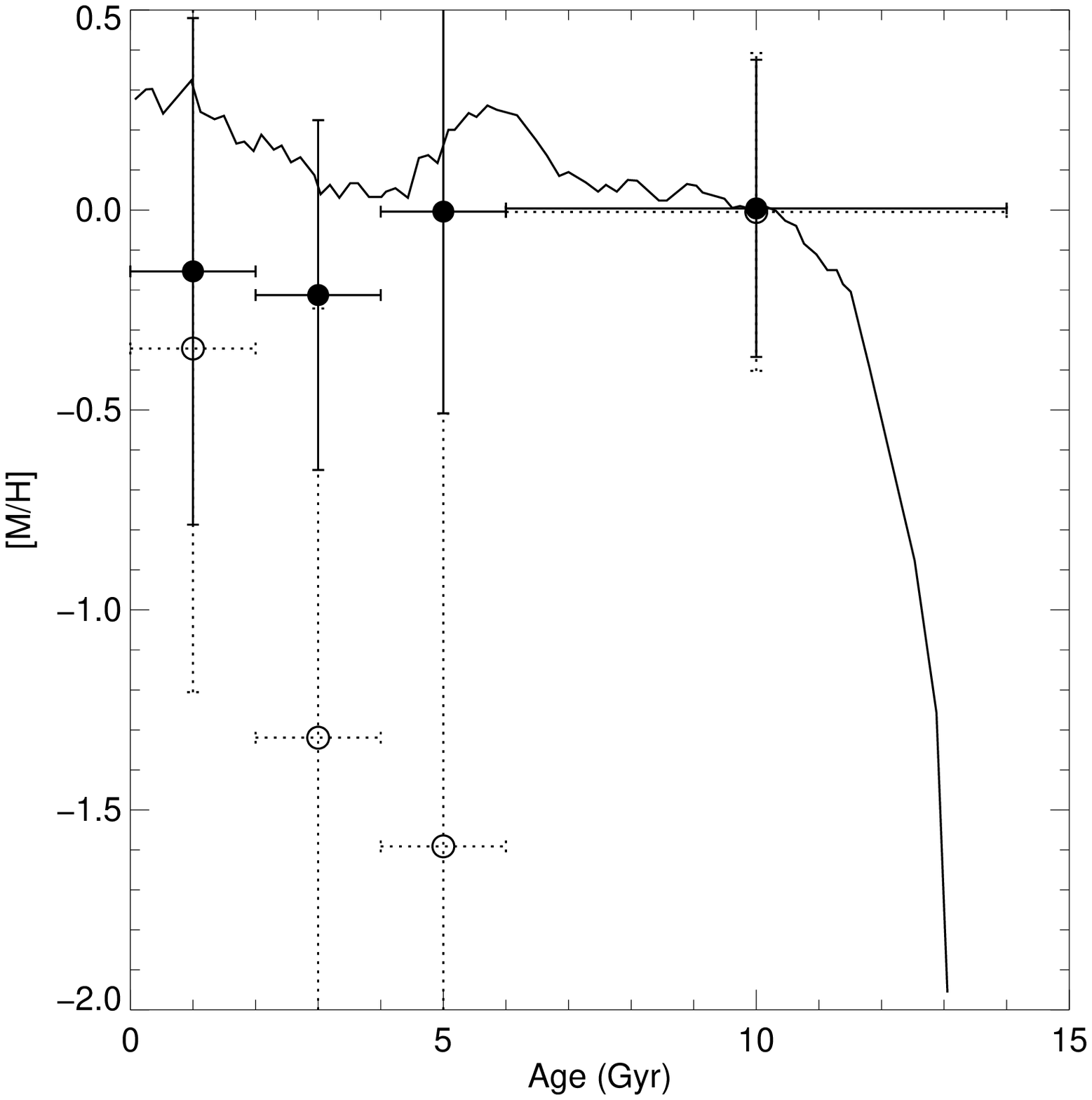}

\end{document}